\documentclass[onecolumn,showpacs,amsmath,prb,10pt]{revtex4}

\usepackage{amsmath}
\usepackage{epsfig}
\usepackage{epsf}
\usepackage{array}
\usepackage{graphicx}

\newcommand{\der}[2]{\frac{d{#1}}{d{#2}}}
\newcommand{\sder}[2]{\frac{d^2{#1}}{d{#2}^2}}
\newcommand{\pder}[2]{\frac{\partial{#1}}{\partial{#2}}}

\def\tr{\textrm{tr}}
\newcommand{\sgn}{\textrm{sgn}}

\def\be{\begin{equation}}
\def\ee{\end{equation}}
\def\bea{\begin{eqnarray}}
\def\eea{\end{eqnarray}}
\def\w{\omega}

\def\e{\epsilon}
\def\ve{\varepsilon}

\def\Tr{\textrm{Tr}}
\def\tr{\textrm{tr}}
\def\t{\theta}

\def\V{{\cal{V}}}

\begin{document}

\title{Nonperturbative interaction effects in the thermodynamics of disordered wires}

\author{D. A. Pesin}
 \affiliation{Department of Physics, University of Washington, Seattle, WA 98195, USA}

\author{A. V. Andreev}
 \affiliation{Department of Physics, University of Washington, Seattle, WA 98195, USA}

\date{\today}

\begin{abstract}
We study nonperturbative interaction corrections to the thermodynamic quantities
of multichannel disordered wires in the presence of the Coulomb interactions.
Within the replica nonlinear $\sigma$-model (NL$\sigma$M) formalism, they arise
from nonperturbative soliton saddle points of the NL$\sigma$M action. The problem
is reduced to evaluating the partition function of a replicated classical one
dimensional Coulomb gas. The state of the latter depends on two parameters: the
number of transverse channels in the wire, $N_{ch}$, and the dimensionless
conductance, $G(L_T)$, of a wire segment of length  equal to the thermal
diffusion length, $L_T$. At relatively high temperatures, $G(L_T) \gtrsim \ln
N_{ch} $,  the gas is dimerized, i.e. consists of bound neutral pairs. At lower
temperatures, $\ln N_{ch} \gtrsim G(L_T) \gtrsim 1$, the pairs overlap and form a
Coulomb plasma. The crossover between the two regimes occurs at a parametrically
large conductance $G(L_T) \sim \ln N_{ch}$, and may be studied independently from
the perturbative effects. Specializing to the high temperature regime, we obtain
the leading nonperturbative correction to the wire heat capacity. Its ratio to
the heat capacity for noninteracting electrons, $C_0$, is  $\delta C/C_0\sim
N_{ch}G^2(L_T)e^{-2G(L_T)}$.
\end{abstract}


\maketitle

\section{Introduction}

The interplay between disorder and electron-electron interactions
in conductors influences their low-temperature properties in an essential
way~\cite{Altaronov,EfrosShklovskii}. Depending on the disorder strength, the
temperature and other system parameters a conductor may be either in the metallic
or in the insulating regime. The manifestations of electron-electron interactions
in the two regimes are quite different. In the insulating regime the charge in a
given localized site is quantized in the units of the electron charge, and charge
discreteness effects dominate the system properties~\cite{EfrosShklovskii}. In
the metallic regime the charge in a given volume of the conductor can change
continuously and charge discreteness effects are small. The two regimes can be
distinguished by the value of the appropriately defined dimensionless conductance
$G$, which is greater than unity in the metallic regime and smaller than unity in
the insulating one. If the system crosses over from the metallic to the
insulating regime due to a change in temperature or disorder strength the Coulomb
blockade effects are expected to gradually grow and become important at $G \sim
1$.

Theoretically the transition between the metallic and the insulating regimes is
typically approached from the metallic side, $G \gg 1$, where electron transport
can be described semiclassically. Therefore the study of incipient charge
discreteness effects in the metallic regime is an important problem in the theory
of disordered conductors. This problem has recently attracted much attention
~\cite{Korshunov,Schoen,Matveev,Wang,Nazarov,Kamenev, Aleiner,Beloborodov2002,Skvortsov,BAL,Altland,MeyerPRB,Andreev,Meyer,AADP_PRL}. In the metallic regime $1/G$ may be used as a
small expansion parameter. For $G\gg 1$ the charge discreteness effects are
exponentially small in $G$, and their analysis requires nonperturbative methods.
To date quantitative studies of nonperturbative interaction effects in the
metallic regime have been limited to granulated systems, or to systems in which
the electron-electron interaction is spatially separated from the disorder. The
present paper is devoted the study of nonperturbative effects in the
thermodynamic properties of homogeneously disordered wires, in which
electron-electron interactions and disorder spatially coexist.

The most promising technique to study this problem is the nonlinear
$\sigma$-model (NL$\sigma$M), either in the replica~\cite{Finkelstein} or
Keldysh~\cite{AK,Nayak} formulation. We use Finkelstein's~\cite{Finkelstein}
replica formulation of the NL$\sigma$M. We show that nonperturbative corrections
to the thermodynamic quantities of the wire depend on two parameters: the number
of channels, $N_{ch}$, in the wire, and the dimensionless conductance, $G(L_T)$,
of the wire segment of length equal to the thermal diffusion length, $L_T$. In
contrast, the perturbative corrections~\cite{Altaronov} are controlled by a
single parameter, $G(L_T)$. For example, the leading perturbation theory
correction to the heat capacity is $\delta C_{PT}/C_0 \sim 1/G(L_T)$, where $C_0$
is the wire heat capacity in the noninteracting electron approximation.

Within the NL$\sigma$M formalism, the nonperturbative effects are described by
soliton saddle points of the NL$\sigma$M action. The spatial extent of the
solitons is given by the thermal diffusion length $L_T$, and their action is
equal to  $G(L_T)$. The nonperturbative contribution to the thermodynamic
quantities is described by the partition function for a gas of these solitons. We
map the problem onto a one dimensional replicated Coulomb gas. At high
temperatures, $G(L_T)\gtrsim \ln N_{ch}$, the Coulomb gas is dimerized, i.e.
consists of widely separated neutral pairs (dimers). In the temperature range
$\ln N_{ch}\gtrsim G(L_T)\gtrsim1$, the dimers are ionized and form a Coulomb
plasma. Since the crossover between the two regimes occurs at a parametrically
large conductance, $G(L_T)\sim \ln N_{ch}$, it can be studied independently from
the perturbative effects. In this paper we specialize to the high temperature
regime, leaving consideration of the crossover to the low temperature one for
future work.

The paper is organized as follows. In Sec.~\ref{section:NLsM} we
describe the NL$\sigma$M for multichannel wires. In
Sec.~\ref{section:Ninfty} we obtain the analytic solution
for the saddle points of the NL$\sigma$M action in the limit of the
infinite number of channels $N_{ch}$,  and evaluate the functional integral over the
fluctuations about the saddle points. In Sec.~\ref{sec:inst_gas} we
obtain the leading nonperturbative correction to the thermodynamic
quantities of the wire for $N_{ch}\gg 1$. In
Sec.~\ref{section:Conclusion} we summarize our results.

\section{Nonlinear $\sigma$-model}\label{section:NLsM}

We consider an infinitely long disordered wire with many transverse channels,
$N_{ch}\gg 1$. The disorder is assumed to be weak, so that the elastic mean free
path $l$ satisfies the condition $k_Fl\gg1$, where $k_F$ is the Fermi wave
number. We consider the temperature $T$ to be smaller than the Thouless energy
for the transverse motion, $E_T\equiv D/d^2$, where $d$ is the transverse wire
dimension, and $D$ is the diffusion constant. In this regime the wire is
described by the one-dimensional $NL\sigma M$.

Thermodynamic properties of the system can be extracted from the averaged over
disorder realizations replicated partition function, $\langle Z^p\rangle=\langle
\Tr e^{-p\frac{\hat{H}}{T}}\rangle$, with $p$ being the number of replicas. We
will be interested in the thermodynamic potential, which can be obtained using
the replica trick:
\begin{equation}\label{replicatrick}
  \langle\Omega\rangle=-T\langle\ln Z\rangle=-T\lim_{p\rightarrow 0}\frac{\langle
  Z^p\rangle-1}{p}.
\end{equation}
 In the diffusive regime the replicated partition function,  $\langle
Z^p\rangle$, has a functional integral representation in terms of NL$\sigma$M,
describing the low-energy physics of the problem. The derivation of the
NL$\sigma$M action  has become a standard procedure~\cite{Finkelstein, Efetov83}.
Therefore, below we only present its final form, suitable for the problem under
consideration. The NL$\sigma$M action is a functional of two fields: the
$Q$-matrix, parameterizing the diffusive degrees of freedom of electron motion,
and electric potential $V$. The former is a Hermitian matrix in the space of
replicas and Matsubara frequencies, whose entries are $4\times 4$ matrices in the
space $S\otimes \mathcal{T}$, given by the product of spin, $S$, and
time-reversal, $\mathcal{T}$, spaces~\cite{Efetov83,Efetov}. The slowly varying
in space electric potential $V_a$ is introduced to treat the the long range part
of the Coulomb interaction in the replica $a$. This part of the Coulomb
interaction  is of particular importance for the consideration below. It cannot
be described by the Fermi-liquid interaction constants.  Since the Fermi-liquid
effects in disordered metals have been studied by Finkelstein~\cite{Finkelstein}
and are not essential for the phenomena discussed in this paper, we ignore them
in order to keep the presentation more transparent. Then the NL$\sigma$M action
can be written as
\begin{subequations}\label{action}
\begin{eqnarray}
\label{action_a}
\langle Z^p \rangle&=&\int {\cal{D}}[Q,V]e^{-S_{Q}-S_C},\\
\label{action_b} S_Q&=&A\frac{\pi\nu}{2}\int dx\textrm{Tr}
\left[\frac{D}{4}(\nabla
Q)^2-(\hat{\varepsilon}+\hat{V})Q\right]\nonumber\\
&&+ A \nu\int d\tau dx\sum_a V^2_a(x,\tau),\\
 \label{action_c}
S_C&=&\!\frac{1}{2}\!\int \!d\tau dx dx'\sum_a
V_a(x,\tau)K(x-x')V_a(x',\tau),
\end{eqnarray}
\end{subequations}
where $\textrm{Tr}$ denotes the trace over the replica, Matsubara and $S\otimes
\mathcal{T}$ spaces, $\nu$ is the density of states per spin at the Fermi level,
and $A$ in the wire cross section area. The matrices $\hat\varepsilon$ and
$\hat{V}$ have the following structure in the replica and $S\otimes \mathcal{T}$
spaces: $\hat\varepsilon=i\delta^{ab} \tau_3
\partial_\tau $, $\hat{V}=\delta^{ab}\tau_0V_a$,
with $\tau_i$'s defined as $\tau_i=t_i\otimes\sigma_0$, where $\sigma_i$, $t_i$ are the Pauli matrices in the $S$ and $\mathcal{T}$ spaces. The term $S_Q$
defined in Eq.~(\ref{action_b}), represents the part of the action that describes
electrons moving in the presence of the auxiliary fields $V_a$, whereas $S_C$,
defined in Eq.~(\ref{action_c}), is the bare Coulomb action. The kernel $K(x-x')$
describes the inverse effective Coulomb interaction in the wire. In particular,
for a homogeneous wire in the absence of a nearby gate its Fourier transform is
$K(q)=1/e^2\ln{\frac{1}{q^2d^2}}$. We also assume that the external magnetic
field is absent. The action~(\ref{action}) constitutes the NL$\sigma$M.

The $Q$-matrix satisfies the nonlinear constraint $Q^2=\openone$. It also
satisfies the charge conjugation condition \cite{Efetov83},
\begin{widetext}
\begin{equation}\label{matrix:chargeconjugation}
Q=CQ^TC^T,\,\,C=\delta^{ab}\delta_{\ve\ve'}\otimes\left(\begin{array}{cccc}
        0&0&0&-1\\
        0&0&1&0\\
        0&-1&0&0\\
        1&0&0&0
        \end{array}\right)\equiv\delta^{ab}\delta_{\ve\ve'}\otimes (t_1\otimes (-i\sigma_2)),
\end{equation}
\end{widetext}
where $a,b$ and $\ve,\ve'$ denote replica and Matsubara indices respectively, and the
superscript $T$ denotes the transposition. In what follows we restrict ourselves to the case of strong spin-orbit scattering. In this case the $Q$-matrix belongs to the symplectic ensemble~\cite{Efetov83}, and its matrix elements are unit matrices in the spin space.

To resolve the nonlinear constraint $Q^2=\openone$ we will use the
exponential parameterization of the $Q$-matrix,
\begin{equation}\label{eq:parameterization}
  Q= e^{iW/2}\Lambda e^{-iW/2},\,\,\,\Lambda^{ab}_{\ve\ve'}=
\delta^{ab}\delta_{\ve\ve'}\tau_0\textrm{sgn}\varepsilon,\,\,\{W,\Lambda\}=0,\,\, W=W^{+},
\end{equation}
where $\{A,B\}$ denotes the anticommutator of $A$ and $B$. The invariance of the $Q$-matrix with respect to the operation of charge conjugation, Eq.~(\ref{matrix:chargeconjugation}), and its hermiticity impose the following matrix structure on the rotation generators, $W$, in the $\mathcal{T}$-space:
\begin{equation}\label{matrix:W}
W^{ab}_{\ve\ve'}=\left(\begin{array}{cc}
        d&c\\
        -c^*&-d^*
        \end{array}\right)^{ab}_{\ve\ve'},\hat{d}^+=\hat{d},\,
        \hat{c}^T=-\hat{c}.
\end{equation}
The fields $(d,c)^{ab}_{\ve\ve'}$ represent the diffuson and cooperon degrees of
freedom respectively, each being a unit matrix in the spin space.

The action in Eq.~(\ref{action}) is characterized by two parameters. The first
is $G(L_T)=4 \pi \hbar\nu D A /L_T$, where $L_T=\sqrt{\hbar D/2\pi T}$ is the thermal diffusion length.
It has the meaning of the dimensionless conductance of the wire segment of
length $L_T$. The other one is the number of transverse channels in the wire,
$N_{ch}=k_F^2A/4\pi$. We consider a multichannel metallic wire, for which both parameters are large. From now on the Planck's constant $\hbar$ is set to unity.

For large $G(L_T)$ we can evaluate the replicated partition function in the
saddle point approximation. In this approximation the partition function is
written as a sum of the contributions arising from all the saddle points:
\begin{equation}\label{eq:s.p.}
\langle Z^p\rangle=\sum_{\substack{saddle\\points}}e^{-S_{s.p.}}\int
{\cal{D}}[\delta Q,\delta V]e^{-\delta S[\delta Q,\delta V]}.
\end{equation}
Here $S_{s.p.}$ denotes the NL$\sigma$M action evaluated at the saddle point, and
$\delta Q,\delta V$ describe fluctuations of the $Q$-matrix and electric
potentials $V_a$ around a particular saddle point. Finally $\delta S[\delta
Q,\delta V]$ denotes the action change due to these fluctuations. In the next
section we discuss the saddle points of the action~(\ref{action}) in the
$G(L_T)=\mathrm{const}$, $N_{ch}\to \infty$  limit, which is referred to below as
the ``$N_{ch}\to \infty$ limit" for brevity.

\section{Saddle points in the $N_{ch}\to \infty$ limit\label{section:Ninfty}}

If the number of channels in the wire is sufficiently large, $e^2\nu A\gg 1$, one
may neglect the Coulomb action, $S_C$, in Eq.~(\ref{action}) when looking for the
saddle points. This corresponds to the charge neutrality limit~\cite{Andreev},
which can be seen by noting that formally such procedure corresponds to the limit
$e\rightarrow\infty$, which clearly enforces electroneutrality. The saddle point
equations in this limit are obtained by minimizing $S_Q$ in Eq.~(\ref{action_a})
with respect to $V_a$ and $Q$, and read
\begin{subequations}\label{eq:saddle}
  \begin{eqnarray}
    \label{eq:saddle_a}
    &&D\nabla(Q\nabla Q)-[\hat{\varepsilon}+\hat{V},Q]=0,\\
    \label{eq:saddle_b}
    &&V_a-\frac{\pi}{4}\textrm{tr}\,Q^{aa}_{\tau\tau}(x)=0,
  \end{eqnarray}
\end{subequations}
where $\textrm{tr}$ is the trace in the $S\otimes \mathcal{T}$ space only.
Equation (\ref{eq:saddle_a}) is the Usadel equation, and Eq.~(\ref{eq:saddle_b})
represents the charge neutrality condition.

By direct substitution one can check that Eqs.~(\ref{eq:saddle}) possess a set of
stationary spatially uniform solutions,
$Q^{ab}_{\ve\ve'}=\delta^{ab}\delta_{\ve\ve'}\tau_0\textrm{sgn}(\varepsilon+2\pi
Tw_a)$, $V_a=2\pi Tw_a$, which are characterized by a set of integer winding
numbers in each replica, $w_a$. All these solutions represent degenerate  minima
of the action~(\ref{action_b}). The sum $4\sum_aw_a \equiv {\cal{W}}$ (the factor
$4$ here arises from the $4\times 4$ matrix structure of $Q^{ab}_{\ve\ve'}$ in
the $S\otimes \mathcal{T}$ space) defines the trace of the $Q$-matrix,
$\textrm{Tr} \, Q=2{\cal{W}}$. The $Q$-matrices corresponding to the minima with
different $w_a$, but the same ${\cal{W}}$ can be transformed into each other via
continuous rotations in the replica and Matsubara spaces,
Eq.~(\ref{eq:parameterization}). Therefore, the NL$\sigma$M action contains
soliton minima in which the $Q$-matrix and the potentials $V_a$ smoothly
interpolate between their values in different uniform minima~\cite{Andreev}. Such
solitons are similar to those first found in Ref.~\cite{Kamenev}.

\subsection{\label{section:soliton}Single soliton solution}

In this section we find an analytic solution to the saddle point equations
(\ref{eq:saddle}) that correspond to a single soliton. To be specific, we
construct a soliton that connects the following degenerate minima: $Q=\Lambda$,
with all the winding numbers $w_a=0$ at $x=-\infty$, and
$Q^{ab}_{\ve\ve'}=\delta^{ab}\delta_{\ve\ve'}\tau_0\textrm{sgn}(\varepsilon+2\pi
Tw_a)$ at $x=\infty$, with $w_{1,2}=\mp 1$, all the other $w_a$ being zero. This
corresponds to a gradual change in the electric potential in replicas $1$ and
$2$, $V_{1,2}$, from zero at negative spatial infinity to $\mp 2\pi T$ at
positive infinity.

For such a soliton the generator $W_0$ parameterizing the saddle point $Q$-matrix
via Eq.~(\ref{eq:parameterization}) corresponds to a rotation between Matsubara
frequencies $\pi T$ in replica 1, and $-\pi T$ in replica 2. In this subspace
$W_0$ has the following structure:
\begin{equation}\label{matrix:saddle_W}
W_0=\left(\begin{array}{cc}
        0&\hat{\lambda}\\
        \hat{\lambda}^+&0
        \end{array}\right),\ \hat{\lambda}=\left(\begin{array}{cc}
                                     \t_de^{i\phi}&\t_ce^{i\chi}\\
                                     -\t_ce^{-i\chi}&-\t_de^{-i\phi}
                                     \end{array}\right),
\end{equation}
where $\t_d,\t_c, \phi$ and $\chi$ are real parameters. In this equation the
matrix element of $W_0$ in the upper-left-corner corresponds to $(W)^{11}_{\pi
T,\pi T}\equiv0$, the one in the upper-right-corner to $(W)^{12}_{\pi T,-\pi
T}=\hat{\lambda}$, and so on. All the other matrix elements of $W_0$ are zero.

Substituting the rotation generator (\ref{matrix:saddle_W}) into
Eq.~(\ref{eq:parameterization}) we obtain the matrix elements of the $Q$-matrix
that participate in the rotation:
 \begin{widetext}
\begin{eqnarray}\label{eq:kink}
&&\left(\begin{array}{c||c}
       Q^{11}_{\pi T,\pi T}&Q^{12}_{\pi T,-\pi T}\\
       \hline\hline
       Q^{21}_{-\pi T,\pi T}&Q^{22}_{-\pi T,-\pi T}
      \end{array}\right)=\nonumber\\
&&\left(\begin{array}{cc||cc}
       \cos{\t_d}\cos{\t_c}&e^{i(\phi+\chi)}\sin{\t_d}\sin{\t_c}&-ie^{i\phi}\sin{\t_d}\cos{\t_c}&-ie^{i\chi}\cos{\t_d}\sin{\t_c}\\
       e^{-i(\phi+\chi)}\sin{\t_d}\sin{\t_c}&\cos{\t_d}\cos{\t_c}&ie^{-i\chi}\cos{\t_d}\sin{\t_c}&ie^{-i\phi}\sin{\t_d}\cos{\t_c}\\
       \hline\hline
       ie^{-i\phi}\sin{\t_d}\cos{\t_c}&-ie^{i\chi}\cos{\t_d}\sin{\t_c}&-\cos{\t_d}\cos{\t_c}&e^{-i(\phi-\chi)}\sin{\t_d}\sin{\t_c}\\
       ie^{-i\chi}\cos{\t_d}\sin{\t_c}&-ie^{i\phi}\sin{\t_d}\cos{\t_c}&e^{i(\phi-\chi)}\sin{\t_d}\sin{\t_c}&-\cos{\t_d}\cos{\t_c}\\
      \end{array}\right).
\end{eqnarray}
\end{widetext}
All the other matrix elements are those of the $\Lambda$-matrix.

The action for such a $Q$-matrix is independent of the angles $\phi$ and $\chi$
and depends only on $(\nabla\phi)^2$ and $(\nabla\chi)^2$ with positive
coefficients. Therefore the action minimum corresponds to coordinate independent
angles $\phi$ and $\chi$. It can be shown that the soliton solutions with the
minimum action correspond to either $\t_d\neq 0,\, \t_c=0$ (diffusonlike
rotation) or $\t_d=0,\, \t_c\neq 0$ (Cooperon-like rotation). In these cases
substitution of Eq.~(\ref{eq:kink}) into (\ref{eq:saddle_b}) gives
$V_{1,2}(x)=\mp \pi T[1-\cos\theta_{d,c}(x)]$ for the diffusonlike and
Cooperon-like rotations respectively. Then Eq.~(\ref{eq:saddle_a}) yields
\begin{equation}\label{eq:sin-gordon}
\nabla^2\theta_{d,c}-\frac{1}{2L^2_T}\sin{2\theta_{d,c}}=0.
\end{equation}
The solution that corresponds to the sought soliton is
\begin{equation}\label{eq:theta}
\theta_{d,c}(x)=2\arctan\left(e^{(x-x_0)/L_T}\right)\equiv
\t_0(x-x_0),
\end{equation}
giving for the electric potentials
\begin{equation}\label{eq:saddleV}
V_{1,2}(x)=\mp V^0(x-x_0)\equiv \mp \pi T\{1+\tanh[(x-x_0)/L_T]\},
\end{equation}
which clearly satisfies $V_{1,2}(x\rightarrow-\infty)=0$ and
$V_{1,2}(x\rightarrow\infty)=\mp2\pi T$. Here $x_0$ denotes  the soliton
position.

Substituting the saddle point values of $Q$ and $V_a$, Eqs.~(\ref{eq:kink})
and~(\ref{eq:saddleV}), into the action~(\ref{action_b}), we obtain the action
for a single soliton
\begin{equation}
  S_0=G(L_T).\nonumber
\end{equation}
We note that this action does not depend on the soliton position $x_0$, and the
angles $\phi$ and $\chi$ in Eq.~(\ref{eq:kink}). However, for the diffusonlike
($\theta_c=0$) soliton the different values of the angle $\chi$ correspond to the
same $Q$-matrix, and similarly different values of $\phi$ correspond to the same
$Q$-matrix for the Cooperon-like ($\theta_d=0$) soliton. Therefore the action for
the fluctuations about the soliton  has only two zero modes. One is associated
with a translation of the soliton (change in $x_0$). The other corresponds to a
rotation of the $Q$-matrix in the replica and Matsubara space caused by a uniform
change in either $\phi$ or $\chi$, depending on whether we consider a
diffusonlike or a Cooperon-like soliton.  The presence of these zero modes needs
to be borne in mind when integrating over the fluctuations about the soliton
configurations.

\subsection{\label{section:Gaussian Fluctuations}Fluctuations around a single
soliton}

In this section we evaluate the single soliton contribution to the replicated
partition function, Eq.~(\ref{eq:s.p.}), in the $N_{ch}\to \infty$ limit. This
requires evaluating the functional integral over the fluctuations of the
$Q$-matrix and the potentials $V_a$ around the single soliton saddle point.

As was explained at the end of Sec.~\ref{section:soliton}, the fluctuation
spectrum has two zero modes. We show below that all the other fluctuations are
massive and integrate over them in the gaussian approximation. The resulting
fluctuation determinant is convergent and is evaluated below. The integration
over the zero modes is reduced to the integration over the soliton position and
the rotation angle.

The translational zero mode represents a simultaneous spatial shift of the saddle
point solution for the $Q$-matrix and the static (zero Matsubara frequency)
component of the potentials $V_a$. In order to simplify the treatment of this
zero mode we first integrate over the latter. This step involves no
approximations since the action (\ref{action_b}) is quadratic in $V_a$. The
resulting action depends only on the nonzero Matsubara components of $V_a$ and on
the $Q$-matrix. In this representations the zero modes involve only the
$Q$-matrix degrees of freedom, whereas all fluctuations of the nonzero Matsubara
components of $V_a$ are massive. Then the single soliton contribution to the
partition function, Eq.~(\ref{eq:s.p.}), in the $N_{ch}\to \infty$ limit can be
written as $\exp[-G(L_T)]\, \Gamma_p$, where $\Gamma_p$ is the functional
integral over the fluctuations about the soliton and is given by
\begin{equation}
\label{eq:Gamma_def}
  \Gamma_p=\alpha^p\int{\cal{D}}[W,\delta V]e^{-S^{(2)}[W,\delta V]}.
\end{equation}
Here the fluctuations of the nonzero Matsubara components of the electric
potential are denoted by $\delta V_a$, the matrix $W$ parameterizes the deviation
of the $Q$-matrix from the saddle point, and $\alpha^p$ is the factor coming from
integration over the static components of $V_a$. We will see later that in order
to obtain the physical observables we will only need to evaluate $\Gamma_p$ at $p
= 0$. Therefore, the  value of $\alpha$ is of no importance. Finally, the
quadratic fluctuation action $S^{(2)}[W,\delta V]$ is obtained by integrating
over the fluctuations of the static component of $V_a$ in Eq.~(\ref{action_b}),
and expanding the resulting action to the second order in $W$. Its  form depends
on the $Q$-matrix parametrization.

In the remainder of this section we show that the fluctuation integral $\Gamma_p$
can be expressed as
\begin{equation}\label{eq:Gamma_result}
    \Gamma_{p}=\alpha^p G(L_T) \Upsilon_p \int  \frac{d x_0}{L_T},
\end{equation}
where $x_0$ is the position of the soliton and $\Upsilon_p$ is a numerical factor
independent of the system parameters. In order to evaluate the thermodynamic
quantities we need only the $p=0$ value of this quantity, which is calculated
below, $\Upsilon \equiv \Upsilon_{p=0}\approx 8$.

In the remainder of the
present section we derive Eq.~(\ref{eq:Gamma_result}). The presentation is organized as follows. In Sec.~\ref{sec:fluct_action} we give
the expression for the fluctuation action. In Sec.~\ref{section:integration_W} we
carry out the integration over the $Q$-matrix fluctuations.
Section~\ref{section:elpotdet} deals with integration over the electric potential
fluctuations.
The reader not interested in the derivation of Eq.~(\ref{eq:Gamma_result}) may
wish to  proceed directly to Sec.~\ref{sec:inst_gas}, where we use it to evaluate
nonperturbative corrections to the thermodynamic quantities.

\subsubsection{Fluctuation action}\label{sec:fluct_action}

We parameterize the deviations of the $Q$-matrix from the saddle point in terms
of the matrix $W$, whose structure is described by Eq.~(\ref{matrix:W}),  as
follows:
\begin{equation}\label{eq:Q_fluctuations}
  Q= e^{iW_0/2} e^{iW/2}\Lambda e^{-iW/2}e^{-iW_0/2}.
\end{equation}
Here  the matrix $W_0$ parameterizes the saddle point $Q$-matrix. For the soliton
described in Sec.~\ref{section:soliton} it is given by
\begin{equation}\label{matrix:W_0}
W_0(x)=\left(\begin{array}{cc}
        0&i\t_0(x)\tau_i\\
        -i\t_0(x)\tau_i&0
        \end{array}\right).
\end{equation}
Here $i=0,1$ corresponds to diffusonlike and Cooperon-like rotations, $\t_0$ is
defined in Eq.~(\ref{eq:theta}), and we set $\phi=\chi=\pi/2$, and $x_0=0$ for
convenience.

In the following we use dimensionless coordinate $\xi=x/L_T$, dimensionless
fermionic Matsubara frequencies, $\epsilon=\ve/2\pi T$, and dimensionless Matsubara components of the electric potential, $\V^\w=\delta V^\w/ 2\pi T$, where $\w$ is an integer defining the bosonic Matsubara frequency, such that the latter is written as $2\pi T\w$. In these variables the quadratic action in
Eq.~(\ref{eq:Gamma_def}) can be written as
\begin{equation}
  S^{(2)}[W,\V]=S_{\V\V}+S_{WW}+S_{W\V}.
\end{equation}
Here $S_{\V\V}$ denotes the part of the action that is quadratic in the
potentials $\V_a$,
\begin{subequations}\label{eq:gaussian fluctuations}
\begin{equation}\label{action_VV}
     S_{\V\V}=\frac{G(L_T)}{2}\sum_{a}\sum_{\w\neq0}\int
  d\xi \V^{\w}_a(\xi)\V^{-\w}_a(\xi'),
\end{equation}
$S_{WW}$ denotes the part of the action that is quadratic in $W$,
\begin{eqnarray}
      \label{action_W}
    S_{WW}&=&\frac{G(L_T)}{16}\int d\xi\left(\sum_{ab}\sum_{\ve>0,\ve'<0}
    \left\{(\e-\e')\textrm{tr}
    W^{ab}_{\ve\ve'}(W^{ab}_{\ve\ve'})^\dagger+\textrm{tr}
    \nabla{W^{ab}_{\ve\ve'}}\nabla{(W^{ab}_{\ve\ve'})^\dagger}\right\}\right.\nonumber\\
    & &+\left\{-\frac{3}{4}\sin^2{\theta_0}+\frac{(1-\cos{\theta_0})}{2}\right\}\sum_{a\ve}
    (\textrm{tr}W^{1a}_{\pi T,\ve}(W^{1a}_{\pi T,\ve})^\dagger
    +\textrm{tr}W^{a2}_{\ve,-\pi T}(W^{a2}_{\ve,-\pi T})^\dagger)\nonumber\\
    & &+\frac{(\cos{\theta_0}-1)}{2}\sum_{a\ve\ve'}\sgn{\ve}
    (\textrm{tr}W^{1a}_{\ve\ve'}(W^{1a}_{\ve\ve'})^\dagger
    -\textrm{tr}W^{2a}_{\ve\ve'}(W^{2a}_{\ve\ve'})^\dagger)\nonumber\\
    & &\left.-\frac{\sin^2{\theta_0}}{4}\left[\tr\left\{(\tau_iW^{12}_{\pi T, -\pi
    T})^2+(\tau_iW^{21}_{-\pi T, \pi T})^2\right\}-\frac{1}{4}\left(\tr\{\tau_i(W^{12}_{\pi T, -\pi
    T}-W^{21}_{-\pi T, \pi T})\}\right)^2\right]\right),
\end{eqnarray}
and $S_{W\V}$ denotes the part of the action that is linear in $W$ and $\V$,
\begin{eqnarray} \label{action_WV}
     S_{W\V}&=&i\frac{G(L_T)}{8}\int d\xi\sum_{a}\sum_{\w\ve}\V_a^{\w}\sgn{\ve}\,\tr
     W^{aa}_{\ve,\ve+2\pi T\w}\nonumber\\
     &&+i\frac{G(L_T)}{8}\int d\xi\sum_{\w>0}\left[\V_1^\w
     \left\{-\left(\cos\frac{\theta_0}{2}-1\right)\tr [W^{11}_{\pi T,2\pi T(\frac{1}{2}-\w)}]^
     \dagger+\sin\frac{\t_0}{2}\tr[\tau_iW^{12}_{2\pi
     T(\frac{1}{2}+\w),-\pi T}]^\dagger
     \right\}\right.\nonumber\\
     &&+\V_1^{-\w}\left\{-\sin\frac{\t_0}{2}\tr[\tau_iW^{12}_{2\pi T(\frac{1}{2}+\w),-\pi
     T}]
     +\left(\cos\frac{\theta_0}{2}-1\right)\tr W^{11}_{\pi T,2\pi
     T(\frac{1}{2}-\w)}\right\}\nonumber\\
     &&+\V_2^\w
     \left\{-\left(\cos\frac{\theta_0}{2}-1\right)\tr [W^{22}_{-2\pi T(\frac{1}{2}-\w),-\pi
     T}]^\dagger
     -\sin\frac{\t_0}{2}\tr[\tau_iW^{12}_{\pi
     T,-2\pi T(\frac{1}{2}+\w)}]^\dagger\right\}\nonumber\\
     &&\left.+\V_2^{-\w}
     \left\{\sin\frac{\t_0}{2}\tr[\tau_iW^{12}_{\pi T,-2\pi T(\frac{1}{2}+\w)}]
     +\left(\cos\frac{\theta_0}{2}-1\right)\tr W^{22}_{-2\pi T(\frac{1}{2}-\w),-\pi
     T}\right\}\right].
\end{eqnarray}
\end{subequations}
Here ``$\dagger$" denotes the Hermitian conjugation in the $S\otimes \mathcal{T}$
space, i.e. corresponds to complex conjugation and transposition within a
$4\times 4$ block, without interchanging replica or Matsubara indices. The diffusonlike soliton corresponds to $\tau_i=\tau_0$, and $\tau_i=\tau_1$ for the Cooperon-like one. To be specific, in what follows we
consider the case of a diffuson-like soliton, i.e. we set $\tau_i=\tau_0$. In the
Cooperon-like case the treatment exactly parallels the one presented below.

Introducing the notation
\begin{equation}\label{definition:GW}
\Gamma_W=\int {\cal{D}}W\exp(-S_{WW}),
\end{equation}
and
\begin{equation}\label{definition:GammaV}
 \Gamma_V =\int {\cal{D}}[\V]
  e^{-S_{\V\V}}\langle e^{-S_{W\V}}\rangle_{_W} =
  \int {\cal{D}}[\V] e^{ -S_{\V\V}+ \frac{1}{2}\langle
  S_{W\V}^2\rangle_{_W}},
\end{equation}
where $\langle \ldots \rangle_{_W}$ denotes the gaussian average with respect to
the action $S_{WW}$, we can write Eq.~(\ref{eq:Gamma_def}) as
\begin{equation}\label{eq:GammaWV_def}
  \Gamma_p= \alpha^p \, \Gamma_W\Gamma_V .
\end{equation}

We evaluate quantities $\Gamma_W$ and $\Gamma_V$ in Sections~\ref{section:integration_W} and~\ref{section:elpotdet}.

\subsubsection{\label{section:integration_W} Integration over $W$}

We now evaluate the functional integral over the fluctuations of the $Q$-matrix,
$\Gamma_W$ in Eq.~(\ref{definition:GW}). Examination of the quadratic action in Eq.~(\ref{action_W}), shows that the
variables $W^{ab}_{\ve\ve'}$ with different replica or Matsubara indices
fluctuate independently.  Moreover, with the exception of $W^{12}_{\pi T,-\pi
T}$, for each $W^{ab}_{\ve\ve'}$ the actions for the diffusons and Cooperons
constituting it are identical. The term containing $W^{12}_{\pi T,-\pi T}$ is
special because it has the same replica and Matsubara indices as the rotation
generator $W_0$ parameterizing the saddle point. The fluctuations of the diffuson and
Cooperon components of $W^{12}_{\pi T,-\pi T}$ are also independent, but their
propagators are different. In particular, we will see that for a
soliton represented by a diffuson-like rotation only diffuson part of
$W^{12}_{\pi T,-\pi T}$ has zero modes, and vice versa for a Cooperon-like
rotation.

In terms of the diffuson and Cooperon variables, see Eq.~(\ref{matrix:W}), the
action~(\ref{action_W}) can be written as
\begin{widetext}
\begin{equation}\label{eq:WW_action}
  S_{WW}=\sum^{'}_{\substack{ab\\\ve>0,\ve'<0}}\int d\xi\left((d^{ab}_{\ve\ve'})^*
  \hat{L}^{ab}_{\ve\ve'}d^{ab}_{\ve\ve'}+(c^{ab}_{\ve\ve'})^*
  \hat{L}^{ab}_{\ve\ve'}c^{ab}_{\ve\ve'}\right)+
  \int d\xi\left(d^*_s\hat{L}_d d_s+c_s^*\hat{L}_c c_s\right),
\end{equation}
\end{widetext}
where the primed sum means that the term with $a=1, b=2, \ve=\pi T$ and
$\ve'=-\pi T$ is excluded, and $(d,c)_s\equiv (d,c)^{12}_{\pi T,-\pi T}$. The
operators $\hat{L}^{ab}_{\ve\ve'}, \hat{L}_{d,c}$ are all of the
Schr\"{o}dinger type and have the form,
\begin{eqnarray}\label{eq:kernerls}
  \hat{L}_{d,c}&=&\frac{G(L_T)}{4}\left(\hat{L}_{\w=1}+u_{d,c}(\xi)\right),\nonumber\\
  \hat{L}^{ab}_{\ve\ve'}&=&\frac{G(L_T)}{4}\left(\hat{L}_{\e-\e'}+U^{ab}_{\ve\ve'}(\xi)\right),
\end{eqnarray}
with the operator $\hat{L}_{\w}$ defined as
\begin{equation}\label{eq:Lw}
  \hat{L}_{\w}=-\frac{d^2}{d\xi^2}+\w,
\end{equation}
with $\w$ and $\e$ being the appropriate dimensionless Matsubara frequencies. The
potentials $u_{d,c}$ for $d_s$, $c_s$ are given by
\begin{eqnarray}\label{potentials_u}
  u_d(\xi)=-\frac{2}{\cosh^2(\xi)},\,\,\,
  u_c(x)=-\frac{1}{\cosh^2(\xi)}.
\end{eqnarray}
The potentials $U^{ab}_{\ve\ve'}$ depend on the replica and Matsubara indices
involved and can be expressed in terms of the following potentials,
\begin{eqnarray}\label{potentials_U}
  v_{1,2}(\xi)=\frac{1}{2}[1\pm\tanh(\xi)],\,\,\,
  u(\xi)=-\frac{3}{4\cosh^2(\xi)}.
\end{eqnarray}
The expressions for the potentials $U^{ab}_{\ve\ve'}$ in terms of $v_{1,2}(\xi)$
and $u(\xi)$ are summarized in Table~\ref{table:1}.
\begin{table*}
\caption{\label{table:1} Potentials $U^{ab}_{\ve\ve'}$ appearing in the operators $\hat{L}^{ab}_{\ve\ve'}$,  Eq.~(\ref{eq:kernerls}). Each entry gives the potential specific to particular replica and Matsubara indices in terms of the potentials $v_{1,2}$ and $u$ defined in Eq.~(\ref{potentials_U}). The
Latin letters ($j,k$) denote replica indices not equal to $1$ or $2$. }
\begin{tabular}{|c|m{40pt}|m{40pt}|m{40pt}|m{40pt}|m{40pt}|m{40pt}|m{40pt}|m{40pt}|m{40pt}|}
\hline
$\ve\ve' \setminus ab $& $jk$ & $1j$ & $2j$ & $j1$ & $j2$ & $11$ & $12$ & $21$ & $22$ \\
\hline
 $\ve>\pi T,\ve'<-\pi T$ & $0$ & $v_2-1$ & $v_1$ & $v_2$ & $v_2-1$ & $0$ &$2v_2-2$ &$2v_1$& $0$\\
\hline
$\ve=\pi T,\ve'<-\pi T$ &0 &$u$ &$v_1$ &$v_2$ &$v_2-1$ &$v_1+u$ &$v_2+u-1$ & $2v_1$&$0$\\
 \hline
$\ve>\pi T,\ve'=-\pi T$ &0 &$v_2-1$ &$v_1$ &$v_2$ &$u$ &$0$ &$v_2+u-1$ & $2v_1$&$v_1+u$\\
 \hline
$\ve=\pi T,\ve'=-\pi T$ &0 &$u$ &$v_1$ &$v_2$ &$u$ &$v_1+u$ & Excluded &$2v_1$ &$v_1+u$\\
 \hline
\end{tabular}
\end{table*}

The operators $\hat{L}^{ab}_{\ve\ve'}$ and $\hat{L}_{c}$ are positive definite.
The operator $\hat{L}_d$, Eq.~(\ref{eq:kernerls}), with the
potential $u_d$, defined in Eq.~(\ref{potentials_u}), has one zero eigenvalue,
with all the other ones being positive and separated by a finite gap. The
integration over the zero modes requires a special consideration. We therefore
defer the integration over the variables $d_s$ in $\Gamma_W$, Eq.~(\ref{definition:GW}), to the end of this
section and begin by integrating over all the other variables first.  To this end
we introduce an auxiliary quantity $\Gamma'_W$ as
\begin{equation}\label{eq:Gamma_prime}
\Gamma_W=\frac{\int {\cal{D}}[d_s]e^{-\int d\xi
d_s^*\hat{L}_dd_s}}{\int {\cal{D}}[d_s]e^{-\frac{G(L_T)}{4}\int d\xi
d_s^*\hat{L}_{\w=1}d_s}}\,\Gamma'_W\equiv \Gamma_d\, \Gamma'_W.
\end{equation}
Calculation of $\Gamma'_W$ reduces to evaluation of gaussian integrals. Since
$(d,c)^{ab}_{\ve\ve'}$ are complex fields, the integration over each of them gives a
factor of an inverse determinant of the corresponding operator in the quadratic
action, Eq.~(\ref{eq:WW_action}), and we obtain the following expression for
$\Gamma'_W$,
\[
  \Gamma'_W=\frac{\alpha^p}{\det\left(\frac{G(L_T)}{4}\hat{L}_{\w=1}\right)\det\hat{L}_{c}}
  \prod^{'}_{\substack{ab\\\ve>0,\ve'<0}}\left(\det\hat{L}^{ab}_{\ve\ve'}\right)^{-2}.
\]
The prime indicates that the product does not include the contribution from
$a=1,b=2,\ve=\pi T, \ve'=-\pi T$. The operators $\hat{L}^{ab}_{\ve\ve'}$ in the expression for $\Gamma'_W$ can be classified according to whether their replica indices correspond to the replicas participating in the soliton rotation.  In particular, for $a,b>2$, the operators $L^{ab}_{\ve\ve'}$ are insensitive to the presence of a soliton. Denoting each of these operators as $\hat{L}^{jk}_{\ve\ve'}$, we see that the product over the replicas with $a,b>2$ contributes a factor
$\left(\prod_{\ve>0,\ve'<0}\det\hat{L}^{jk}_{\ve\ve'}\right)^{-(p-2)^2}$ to the
fluctuation determinant. Analogously, for $a=1,2$ and $b>2$, we have $p-2$
identical operators $L^{ab}_{\ve\ve'}$ for each of $a=1$ and  $a=2$, which we denote as $\hat{L}^{1j}_{\ve\ve'}$ and  $\hat{L}^{2j}_{\ve\ve'}$ respectively. Finally,
there are $p-2$ equal operators for $a>2$ and each of $b=1$ and $b=2$, denoted as $\hat{L}^{j1}_{\ve\ve'}$ and  $\hat{L}^{j2}_{\ve\ve'}$. Using these observations we rewrite the previous equation as
\begin{widetext}
\begin{eqnarray}
  \Gamma'_W&=&\frac{\alpha^p}{\det\left(\frac{G(L_T)}{4}\hat{L}_{\w=1}\right)\det\hat{L}_{c}}
  \left(\prod^{'}_{\substack{\ve>0,\ve'<0}}\det\hat{L}^{12}_{\ve\ve'}\det\hat{L}^{21}_{\ve\ve'}
  \det\hat{L}^{11}_{\ve\ve'}\det\hat{L}^{22}_{\ve\ve'}\right)^{-1}\nonumber\\
  &&\times\left(\prod_{\ve>0,\ve'<0}\det\hat{L}^{jk}_{\ve\ve'}\right)^{-(p-2)^2}
  \left(\prod_{\ve>0,\ve'<0}\det\hat{L}^{1j}_{\ve\ve'}\det\hat{L}^{2j}_{\ve\ve'}
  \det\hat{L}^{j1}_{\ve\ve'}\det\hat{L}^{j2}_{\ve\ve'}\right)^{-(p-2)}\nonumber
  \end{eqnarray}
  \end{widetext}
In the above expression the prime means that $\det{\hat{L}^{12}_{\pi
T,-\pi T}}$ is excluded from the product. To compute the thermodynamic quantities
we will need only the value of $\Gamma'_W$ at $p=0$, for which we use the same
notation,
\begin{widetext}
\begin{eqnarray}\label{eq:replicadeterminant}
  \Gamma'_W&=&\frac{1}{\det\left(\frac{G(L_T)}{4}\hat{L}_{\w=1}\right)\det\hat{L}_{c}}
  \left(\prod^{'}_{\substack{\ve>0,\ve'<0}}\det\hat{L}^{12}_{\ve\ve'}\det\hat{L}^{21}_{\ve\ve'}
  \det\hat{L}^{11}_{\ve\ve'}\det\hat{L}^{22}_{\ve\ve'}\left[\det\hat{L}^{jk}_{\ve\ve'}\right]^4\right)^{-1}
  \nonumber\\
  &&\times\left(\prod_{\ve>0,\ve'<0}\det\hat{L}^{1j}_{\ve\ve'}\det\hat{L}^{2j}_{\ve\ve'}
  \det\hat{L}^{j1}_{\ve\ve'}\det\hat{L}^{j2}_{\ve\ve'}\right)^{2}.
  \end{eqnarray}
\end{widetext}

Using Eq.~(\ref{action_W}), the
definitions~(\ref{eq:kernerls})-(\ref{potentials_U}), and the identity
$\ln\det\hat{O}=\tr\ln\hat{O}$ we can write
\begin{widetext}
for~(\ref{eq:replicadeterminant})
\begin{eqnarray}\label{eq:fluctuationdeterminant}
  \ln
  \Gamma'_W&=&2\sum^\infty_{\w=1}\left(4\w\tr_\xi\ln\frac{(L_\w+v_1)(L_\w+v_2)}{L_\w(L_\w+1)}
  -\w\tr_\xi\ln\frac{(L_\w+2v_1)(L_\w+2v_2)}{L_\w(L_\w+2)}+
  4\tr_\xi\ln\frac{L_\w+u}{L_\w}\right.\nonumber\\
  &&\left.-2\tr_\xi\ln\frac{(L_\w+v_1+u)(L_\w+v_2+u)}{L_\w(L_\w+1)}\right)
  -\tr_\xi\ln\frac{L_{\w=1}+u_c}{L_{\w=1}},
\end{eqnarray}
\end{widetext}
where $\tr_\xi$ denotes the trace in the coordinate space, $\tr_\xi\hat{O}=\int
d\xi O(\xi,\xi)$. The terms in Eq.~(\ref{eq:fluctuationdeterminant}) are grouped in
such a way that each is finite at a given $\w$, i.e. does not diverge with the length of the system.

 In Appendix~\ref{section:expression for det} it is shown that
 each term in Eq.~(\ref{eq:fluctuationdeterminant}) can be evaluated using the formula
 \begin{equation}\label{eq:exprgamma_h}
  \tr_\xi\ln\frac{(L_\w+U_1)(L_\w+U_2)}{L_\w(L_\w+h)}=\ln tt'
  =\ln \sqrt{\frac{\w+h}{\w}}t^2,
\end{equation}
where the potentials $U_1(\xi)$ and $U_2(\xi)$ satisfy $U_1(\xi)=U_2(-\xi)$,
$U_1(-\infty)=0$,  $U_1(\infty)= h$ (for the potentials from
Eq.~(\ref{eq:fluctuationdeterminant}) the parameter $h$ takes on the values 0, 1 or 2). The
quantities $t,t'$ describe the $\xi \to \infty$ asymptotics of the two
independent solutions of the equation
 \begin{equation}\label{eq:psi}
   \left[\hat{L}_\w+U_1(\xi)\right]\psi=0.
 \end{equation}
Namely, if we find the two solutions, $\psi_{1,2}$, whose asymptotics at $\xi \to \pm \infty$ are given by
$\psi_1(\xi \to -\infty) \approx \exp(\sqrt{\w }\, \xi)$ and $\psi_2(\xi \to
+\infty) \approx \exp(-\sqrt{\w +h} \, \xi)$, the parameters $t$ and $t'$  are
given by the coefficients in front of the growing exponentials in the asymptotics
of these solutions at the opposite infinities,
\begin{widetext}
\begin{equation}\label{asymptotics}
  \psi_1(\xi)\approx t\exp (\sqrt{\w+h}\, \xi), \,\,\,\,\xi\to\infty\, ; \qquad
\psi_2(\xi)\approx t'\exp( -\sqrt{\w}\, \xi) , \,\,\,\, \xi \to -\infty.
\end{equation}
\end{widetext}
The last equality in Eq.~(\ref{eq:exprgamma_h}) holds since
$\sqrt{\w+h}t=\sqrt{\w}t'$, see Appendix~\ref{section:expression for
det} for details. The case of a potential vanishing
at spatial infinities is recovered from Eq.~(\ref{eq:exprgamma_h}) by
setting $h=0$, $U_1=U_2= U$, $t=t'$:
\begin{equation}\label{eq:exprgamma_0}
  \tr_\xi\ln\frac{L_\w+U}{L_\w}=\ln t.
\end{equation}

In order to find the parameters  $t$ and $t'$ corresponding to the potentials in Eq.~(\ref{eq:fluctuationdeterminant}), we note that for each potential Eq.~(\ref{eq:psi}) has the general form
\begin{equation}\label{eq:schroedingerequation}
  -\frac{d^2\psi}{d\xi^2}+\left[\w-\frac{\alpha}{\cosh^2\xi}+\frac{\beta}{2}(1+\tanh
  \xi)\right]\psi=0,
\end{equation}
where the values of the parameters $\alpha, \beta$ depend on the specific
potential. For example, the potentials $u(\xi)$ and $v_1(\xi)$ in Eq.~(\ref{potentials_U}) correspond to
$\alpha=3/4,\beta=0$ and $\alpha=0, \beta=1$ respectively.

If one introduces the variable $z=(1+\tanh \xi)/2$, and
$y(z)=z^{-\sqrt{\w}/2}(1-z)^{-\sqrt{\w+\beta}/2}\psi(z)$, the above equation
reduces to the hypergeometric equation for $y(z)$:
\begin{equation}
  z(1-z)\sder{y}{z}+\left[c-(a+b+1)z\right]\der{y}{z}-aby=0,
\end{equation}
where the parameters $a$, $b$, and $c$ are given by the following expressions:
\begin{eqnarray}\label{constants}
  c&=&1+\sqrt{\w},\nonumber\\
  a&=&\frac{1}{2}(1+\sqrt{\w}+\sqrt{\w+\beta}-\sqrt{1+4\alpha}),\nonumber\\
  b&=&\frac{1}{2}(1+\sqrt{\w}+\sqrt{\w+\beta}+\sqrt{1+4\alpha}).
\end{eqnarray}

Using the properties of the hypergeometric functions $F(a,b,c,z)$~\cite{mathref}
and switching back to the original variable $\xi=\textrm{arctanh}(2z-1)$ it is
easy to show that the two independent solutions $\psi_{1,2}$ of
Eq.~(\ref{eq:schroedingerequation}) satisfying the desired asymptotics,
$\psi_1(\xi \to -\infty) \to \exp(\sqrt{\w} \xi)$ and $\psi_2(\xi \to \infty) \to
\exp(-\sqrt{\w +h} \xi)$, are given by
\begin{eqnarray}\label{schroedingersolutions}
\psi_1(z)&=&z^{\frac{\sqrt{\w}}{2}}(1-z)^{\frac{\sqrt{\w+\beta}}{2}}F(a,b,c,z),\nonumber\\
\psi_2(z)&=&z^{\frac{\sqrt{\w}}{2}}(1-z)^{\frac{\sqrt{\w+\beta}}{2}}F(a,b,a+b-c+1,1-z).
\end{eqnarray}
The asymptotic behavior of $\psi_1(\xi)$ at $\xi \to +\infty$ is
\begin{eqnarray}\label{psi_boundary}
  \psi_1(\xi\to +\infty )\approx \frac{\Gamma(c)\Gamma(a+b-c)}{\Gamma(a)\Gamma(b)}\, e^{\sqrt{\w+\beta}
  \xi},
\end{eqnarray}
where $\Gamma(x)$ is the Euler gamma function. Comparing Eq.~(\ref{psi_boundary})
with Eq.~(\ref{asymptotics}), we find that the value of the coefficient $t$
entering Eq.~(\ref{eq:exprgamma_h}) is given by
\begin{equation}\label{eq:t}
  t=\frac{\Gamma(c)\Gamma(a+b-c)}{\Gamma(a)\Gamma(b)},
\end{equation}
with $a,b,c$ defined in Eq.~(\ref{constants}). Using
Eqs.~(\ref{eq:exprgamma_h}),~(\ref{eq:exprgamma_0}) and~(\ref{eq:t}) and the
identity $\Gamma(x+1)=x\Gamma(x)$ we obtain for $\Gamma'_W$,
Eq.~(\ref{eq:fluctuationdeterminant}):
\begin{widetext}
\begin{eqnarray}\label{GW}
  \ln
  \Gamma'_W&=&2\sum^\infty_{\w=1}\left(4\w\ln\left[\frac{4\sqrt{\w}\sqrt{\w+1}}
  {(\sqrt{\w}+\sqrt{\w+1})^2}\frac{\Gamma^2(\sqrt{\w})\Gamma^2(\sqrt{\w+1})}
  {\Gamma^4(\sqrt{\w}/2+\sqrt{\w+1}/2)}\right]
  -\w\ln\left[\frac{4\sqrt{\w}\sqrt{\w+2}}
  {(\sqrt{\w}+\sqrt{\w+2})^2}\frac{\Gamma^2(\sqrt{\w})\Gamma^2(\sqrt{\w+2})}
  {\Gamma^4(\sqrt{\w}/2+\sqrt{\w+2}/2)}\right]\right.\nonumber\\
  &&\left.-2\ln\left[\frac{16\sqrt{\w}\sqrt{\w+1}}
  {((\sqrt{\w}+\sqrt{\w+1})^2-1)^2}\frac{\Gamma^2(\sqrt{\w})\Gamma^2(\sqrt{\w+1})}
  {\Gamma^4(\sqrt{\w}/2+\sqrt{\w+1}/2-1/2)}\right]+4\ln\left[\frac{\sqrt{\w}}
  {(\w-1/4)}\frac{\Gamma^2(\sqrt{\w})}
  {\Gamma^2(\sqrt{\w}-1/2)}\right]\right)\nonumber\\
  &&+\ln\left(\Gamma\left(\frac{3-\sqrt{5}}{2}\right)\Gamma\left(\frac{3+\sqrt{5}}{2}\right)\right)\approx
  \ln 0.08.
\end{eqnarray}
\end{widetext}
One can check that the sum over $\w$ above converges, since the summand behaves
like $\w^{-3/2}$ for large $\w$. The last equality was obtained by performing the
summation numerically.

We now complete the evaluation of $\Gamma_W$ by computing the functional integral
$\Gamma_d$, defined in Eq.~(\ref{eq:Gamma_prime}). The operator $\hat{L}_d$
defined by Eqs.~(\ref{eq:kernerls}) and (\ref{potentials_u}) has one zero
eigenvalue. The corresponding eigenfunction is $1/\cosh \xi$. The fluctuations of
$\Re d_s$ and $\Im d_s$ in the numerator of Eq.~(\ref{eq:Gamma_prime}) along this
mode correspond to the rotational and translational zero modes of the soliton
discussed at the end of Sec.~\ref{section:soliton}.

Indeed, in the parametrization (\ref{eq:Q_fluctuations}) a soliton displacement,
$Q_0(\xi) \to Q_0(\xi -\xi_0)$, by a small amount $\xi_0 =x_0/L_T$ is described
by the generator $W_{\xi_0}$ that can be obtained from the condition
\begin{equation}
  \delta Q\approx -ie^{iW_0/2}\Lambda W_{\xi_0}e^{-iW_0/2}=-\pder{Q_0}{\xi} \,
  \xi_0, \nonumber
\end{equation}
where $Q_0$ is  given by Eq.~(\ref{eq:Q_fluctuations}) with $W=0$ and $W_0$ from
Eq.~(\ref{matrix:W_0}) with $i=0$. From this equation it follows that $W_{\xi_0}$ has the same structure as $W_0$, Eq.~(\ref{matrix:saddle_W}), with matrix $\hat{\lambda}$ replaced by $\left(W_{\xi_0}\right)^{12}_{\pi T,-\pi T}$ defined as
\begin{equation}
  \left(W_{\xi_0}\right)^{12}_{\pi T,-\pi T}=- i\tau_0\der{\t_0 (\xi)}{\xi}
  \xi_0.\nonumber
\end{equation}
Comparing this expression with Eq.~(\ref{matrix:W}), we see that the soliton
translation corresponds to the diffuson fluctuation of the form $d_s(\xi)= -i
\der{\t_0(\xi)}{\xi}\, \xi_0=-(i/\cosh{\xi}) \, \xi_0$. Along the same
lines of reasoning it can be shown that the soliton rotation by the angle $
\phi_0$, $\phi \to \pi/2 + \phi_0$ in Eq.~(\ref{eq:kink}),
 corresponds to $ d_s (\xi)=
(1/\cosh{\xi}) \, \phi_0$, and represents the other zero mode.

We separate the functional integral over $d_s$ in the numerator of
Eq.~(\ref{eq:Gamma_prime}) into a product of integrals over the zero and massive
modes:
\begin{widetext}
\begin{equation}
\Gamma_d= \frac{ J\int d\xi_0\int d\phi_0\int
{\cal{D}}[\tilde{d_s}]e^{-\frac{G(L_T)}{4}\int d\xi
\tilde{d_s}^*(\hat{L}_{\w=1}+u_d)\tilde{d_s}} }{ \int
{\cal{D}}[d_s]e^{-\frac{G(L_T)}{4}\int d\xi d_s^*\hat{L}_{\w=1}d_s} } , \nonumber
\end{equation}
\end{widetext}
where $\tilde{d_s}$ contains the massive modes only, and $J$ denotes the Jacobian
for the change of variables $\{d_s\}\rightarrow \{\tilde{d}_s,\xi_0,\phi_0\}$.
The product of the Jacobian $J$ and the ratio of the functional integrals in this expression can
be evaluated using the following trick. We introduce a regularized ratio
$\Gamma_d(\eta)$ of the functional integrals over $d_s$ in the last equation by
infinitesimally shifting the frequency $\w$ from unity, $\w\rightarrow 1+\eta$,
where $\eta$ is positive. As a result the zero modes acquire a finite mass and
$\Gamma_d(\eta)$ can be written as
\begin{equation} \label{eq:Gamma_d_eta}
  \Gamma_d(\eta) \equiv
\frac{ J\int d\xi_0\int d\phi_0 \exp\left[-\frac{\eta G(L_T)}{4}\int
\frac{d\xi}{\cosh^2\xi}
  (\xi_0^2+\phi_0^2)\right]\int
{\cal{D}}[\tilde{d_s}]e^{-\frac{G(L_T)}{4}\int d\xi
\tilde{d_s}^*(\hat{L}_{\w=1}+u_d)\tilde{d_s}} }{ \int
{\cal{D}}[d_s]e^{-\frac{G(L_T)}{4}\int d\xi d_s^*\hat{L}_{\w=1}d_s} } .
\end{equation}
On the other hand this ratio of gaussian integrals can be calculated using
Eq.~(\ref{eq:exprgamma_0}), (\ref{eq:t}), and (\ref{constants}). In the limit of
$\eta \to 0$ we obtain
\begin{widetext}
\begin{equation}\label{eq:Gamma_d_modes}
  \Gamma_d(\eta)=  \mathrm{det} \frac{\hat{L}_{\w=1+\eta}}{\hat{L}_{\w=1+\eta}+u_d}
=\frac{\Gamma(\sqrt{1+\eta}-1)\Gamma(\sqrt{1+\eta}+2)}
  {\Gamma(\sqrt{1+\eta})\Gamma(\sqrt{1+\eta}+1)}\approx\frac{ 4}{\eta}.
\end{equation}
\end{widetext}
To arrive at this expression set $\w =1+\eta$  in Eq.~(\ref{constants}) and used
the fact that the potential $u_d$ corresponds to $\alpha=2$ and $\beta=0$.
Integrating over $\xi_0$  and $\phi_0$ in Eq.~(\ref{eq:Gamma_d_eta}) and
comparing the result with Eq.~(\ref{eq:Gamma_d_modes}) we conclude that
$\Gamma_d$ can be written as
\[
\Gamma_d= \frac{2\, G(L_T)}{\pi} \int d\phi_0 \int d\xi_0.
\]
Substituting this expression into Eq.~(\ref{eq:Gamma_prime}) and integrating over
$\phi_0$ we obtain the following expression for $\Gamma_W$,
\begin{equation}\label{eq:gamma_W_result}
   \Gamma_W=4\Gamma'_W\, G(L_T)\int d\xi_0,
\end{equation}
with $\Gamma'_W$ given by Eq.~(\ref{GW}).

\subsubsection{\label{section:elpotdet}Integration over the electric potential
fluctuations}

We now turn to the evaluation of the functional integral over the potential
fluctuations, $\Gamma_V$ in Eq.~(\ref{definition:GammaV}). The action for the
potential fluctuations is obtained by evaluating the gaussian average, $\langle
S_{W\V}^2 \rangle_{_W}$,  in Eq.~(\ref{definition:GammaV}) with respect to the
action $S_{WW}$ in Eq.~(\ref{action_W}). The result of this tedious, but
straightforward calculation can be expressed in the form
\begin{eqnarray}\label{eq:polarization_soliton}
  S_{\V \V}-\frac{1}{2}\langle S_{W\V}^2 \rangle_{_W} &=& \frac{G(L_T)}{2}\sum_{\w\neq0}\int
  d\xi d\xi' \left[ \sum_{a =1}^p\V^\w_a(\xi)\Pi^\w_0(\xi-\xi')\V^{-\w}_a(\xi')
  -
  \sum_{a=1,2}\V^\w_a(\xi)\delta \Pi^\w(\xi,\xi')\V^{-\w}_a(\xi') \right] .
\end{eqnarray}
Here, in the $p-2$ replicas not participating in the soliton rotation, the
dimensionless polarization operator $\Pi^{\omega}_0(\xi-\xi')$ is given by the
usual expression,
\begin{eqnarray}\label{definition:Pi_0}
\Pi^{\omega}_0(\xi-\xi')&=&\int\frac{dq}{2\pi}e^{iq(\xi-\xi')}\frac{q^2}{|\omega|+q^2},
\end{eqnarray}
and in the remaining two replicas ($a=1,2$) the dimensionless polarization
operators acquire a correction $\delta\Pi^\w(\xi,\xi')$ due to the presence of a
soliton,
\begin{widetext}
\begin{eqnarray}\label{definition:Pi_s}
\delta\Pi^\w(\xi,\xi')&=&\cos{\frac{\theta(\xi)}{2}}G^\w_1(\xi,\xi')\cos{\frac{\theta(\xi')}{2}}+
  \sin{\frac{\theta(\xi)}{2}}G^\w_2(\xi,\xi')\sin{\frac{\theta(\xi')}{2}}-G^\w_0(\xi-\xi').
\end{eqnarray}
\end{widetext}
In the last equation we introduced the following Green's functions:
\begin{eqnarray}
G^{\omega}_0(\xi-\xi')&=&\hat{L}^{-1}_{|\w|}\nonumber\\
G^{\omega}_{1,2}(\xi,\xi')&=&(\hat{L}_{|\w|}+v_{1,2}+u)^{-1},
\end{eqnarray}
where the operator $L_\w$ and the potentials $v_1(\xi),v_2(\xi)$, and $u(\xi)$
are defined in Eqs.~(\ref{eq:Lw}) and~(\ref{potentials_U}).

We note that the polarization operator in the presence of the soliton,
Eq.~(\ref{eq:polarization_soliton}), is diagonal in Matsubara frequencies. This
is a consequence of the fact that the soliton saddle point is static. We also
note that no inter-replica couplings between the potential fluctuations are
generated.

As an important consistency check, let us prove that
\begin{equation}
\int^{\infty}_{-\infty}d\xi'\Pi^{\w\neq 0}_s(\xi,\xi')=0,
\end{equation}
which must hold due to particle number conservation. The polarization operator $\Pi^\w_0$
automatically satisfies this property, as its Fourier transform is
proportional to $q$. To prove that $\delta \Pi^\w$ satisfies the
same condition, we note that
$(-\sder{}{\xi}+v_1+u)\cos{\frac{\theta_0}{2}}=0$, and
$(-\sder{}{\xi}+v_2+u)\sin{\frac{\theta_0}{2}}=0$. Thus we can write
\begin{widetext}
\begin{eqnarray}\label{eq:gauge invariance}
\int^{\infty}_{-\infty}d\xi'\delta\Pi^\w(\xi,\xi')&=&
\int^{\infty}_{-\infty}d\xi'\cos{\frac{\theta_0(\xi)}{2}}
\,G^{\omega}_1(\xi,\xi')\cos{\frac{\theta_0(\xi')}{2}}+
\int^{\infty}_{-\infty}d\xi'\sin{\frac{\theta_0(\xi)}{2}}\,G^{\omega}_2(\xi,\xi')
\sin{\frac{\theta_0(\xi')}{2}}\nonumber\\
&&-\int^{\infty}_{-\infty}d\xi'G^{\w}_0(\xi-\xi')=
\frac{1}{|\w|}\cos^2{\frac{\t_0}{2}}+\frac{1}{|\w|}\sin^2{\frac{\t_0}{2}}-\frac{1}{|\w|}=0,
\end{eqnarray}
\end{widetext}
as expected.

Performing the gaussian integral over $\V$ in Eq.~(\ref{definition:GammaV}) and
taking the number of replicas $p$ to zero we obtain
\begin{widetext}
\begin{eqnarray}\label{eq:detV}
\Gamma_V=&\prod_{\w>0}\frac{\det^2\Pi^\w_0}{\det^2(\Pi^\w_0-\delta \Pi^\w)}=
e^{-2\sum_{\omega>0}\textrm{tr}_\xi\ln\left[1-\delta\Pi^\w(\Pi^\w_0)^{-1}\right]}.
\end{eqnarray}
\end{widetext}
Due to the complicated form of $\delta\Pi^\w$ evaluation of this quantity
explicitly is a daunting task. In particular, the method of the previous section
does not apply here because the polarization operators are not represented by
Schr\"{o}dinger-type operators. However, we note that the dimensionless
polarization operators $\Pi^\w_0$ and $\Pi^\w_0- \delta \Pi^\w$ are independent
of the system parameters. Provided the sum over $\w$ in Eq.~(\ref{eq:detV}) converges, it is clear that $\Gamma_V$ is a parameter-independent numerical factor. Below we prove that the sum over $\w$ in the exponent of Eq.~(\ref{eq:detV}) does converge, and  evaluate $\Gamma_V$ numerically.

To this end we obtain the large-$\w$ asymptotics of the summand in
Eq.~(\ref{eq:detV}). This can be done by expanding the logarithm
in Eq.~(\ref{eq:detV}) to the first order in $\delta\Pi^\omega (\Pi^\omega_0)^{-1}$,
\begin{widetext}

\begin{equation}\label{eq:trace}
 \textrm{tr}_\xi\ln\left[1-\delta\Pi^\w(\Pi^\w_0)^{-1}\right]\approx
 -\tr_\xi\delta\Pi^\w(\Pi^\w_0)^{-1}=-\int
  d\xi\,d\xi'\delta\Pi^\w(\xi,\xi')(\Pi^\w_{0})^{-1}_{\xi'-\xi}.
\end{equation}
\end{widetext}
Using the Fourier transform of $\Pi^\w_0$ from Eq.~(\ref{definition:Pi_0}), the above trace can be written as
\begin{widetext}
\begin{equation}\label{eq:trace_fouried}
  \tr_\xi\delta\Pi^\w(\Pi^\w_0)^{-1}=
  \int\frac{dq}{2\pi}\frac{\w+q^2}{q^2}\int
  d\xi\,d\xi'e^{-iq\xi}\delta\Pi^\w(\xi,\xi')e^{iq\xi'}.
\end{equation}
\end{widetext}
We note that each of the two terms in the expression for $\delta \Pi^\w$,
Eq.~(\ref{definition:Pi_s}), can be written as
$\psi_0(\xi)G(\xi,\xi')\psi_0(\xi')$, where $G(\xi,\xi')$ is the resolvent of of the operator $(\w-\sder{}{\xi}+U)$, and $\psi_0$ is the zero mode of $(-\sder{}{\xi}+U)$. The
phase factors in the last integral in Eq.~(\ref{eq:trace_fouried}) can be
interpreted as a gauge transformation of the Green's function
$\tilde{G}(\xi,\xi')=e^{-iq\xi}G(\xi,\xi')e^{iq\xi'}=(\w+(\frac{1}{i}\der{}{\xi}+q)^2+U)^{-1}$.
Therefore, the integral can be written as
\begin{widetext}
\begin{eqnarray}\label{eq:gauge transform}
\int
  d\xi\,d\xi'e^{-iq\xi}\delta\Pi^\w(\xi,\xi')e^{iq\xi'}&=&\int d\xi d\xi'
  \cos{\frac{\theta_0(\xi)}{2}}
\,\left(\frac{1}{\w+q^2+2q(\frac{1}{i}\der{}{\xi})-\sder{}{\xi}+v_1+u}\right)_{\xi,\xi'}
\cos{\frac{\theta_0(\xi')}{2}}\nonumber\\
&&+\int d\xi d\xi'\sin{\frac{\theta_0(\xi)}{2}}\,
\left(\frac{1}{\w+q^2+2q(\frac{1}{i}\der{}{\xi})-\sder{}{\xi}+v_2+u}\right)_{\xi,\xi'}
\sin{\frac{\theta_0(\xi')}{2}}\nonumber\\
&&-\int d\xi\frac{1}{\w+q^2}.
\end{eqnarray}
\end{widetext}
Then we expand each of the first two kernels in the r.h.s. of Eq.~(\ref{eq:gauge
transform}) in powers of
$\left[-\sder{}{\xi}+v_{1,2}+u+2q(\frac{1}{i}\der{}{\xi})\right]/(\w+q^2)$ to the
order that gives first nonvanishing contribution to the entire integral. The zeroth
order term vanishes in the same way as it happened in Eq.~(\ref{eq:gauge
invariance}), and so does the first order one being proportional to
$\cos{\frac{\theta_0}{2}}\der{}{\xi}\cos{\frac{\theta_0}{2}}
+\sin{\frac{\theta_0}{2}}\der{}{\xi}\sin{\frac{\theta_0}{2}}\equiv\der{}{\xi}(1/2)=0$.
Therefore, expansion to the second order gives the first nonzero contribution,
and we arrive at
\begin{widetext}
\begin{eqnarray}\label{eq:tracefinal}
  \tr_\xi\delta\Pi^\w(K+\Pi^\w_0)^{-1}&\approx&\int \frac{dq}{2\pi}
  \frac{\w+q^2}{q^2}
  (-4)\frac{q^2}{(\w+q^2)^3}\int
  d\xi\left\{\cos{\frac{\theta_0}{2}}\sder{}{\xi}\cos{\frac{\theta_0}{2}}+
  \sin{\frac{\theta_0}{2}}\sder{}{\xi}\sin{\frac{\theta_0}{2}}\right\}\nonumber\\
  &\approx&\frac{1}{2\w^{3/2}}.
\end{eqnarray}
\end{widetext}
Eq.~(\ref{eq:tracefinal}) proves convergence of the sum over
frequencies in Eq.~(\ref{eq:detV}). Therefore we do not have to introduce any
additional regulators.

We can calculate $\Gamma_V$ numerically by expanding the logarithm  in Eq.~(\ref{eq:detV}) v in $\delta\Pi^\omega (\Pi^\omega_0)^{-1}$ and calculating the corresponding traces. The explicit calculation shows that expansion to the third order yields a precision better than a percent, which is sufficient for our purposes. Proceeding this way we obtain $\Gamma_V\approx 24$. Combining $\Gamma_V$ with $\Gamma_W$, expressed via $\Gamma'_W$ and $\Gamma_d$  calculated in Eqs.~(\ref{GW}) and (\ref{eq:gamma_W_result}), we obtain the final expression for $\Upsilon$, determining the fluctuation integral $\Gamma_{p=0}$, Eq.~(\ref{eq:Gamma_result}), needed to calculate the thermodynamics quantities:
\begin{equation}\label{eq:upsilon}
  \Upsilon=4\Gamma'_W\Gamma_V\approx 8.
\end{equation}

\section{Nonperturbative corrections to the thermodynamic quantities}
\label{sec:inst_gas}

In the previous section we found the soliton saddle points and showed that the
functional integral over the fluctuations about a single soliton configuration
can be expressed in terms of the integral over the soliton position,
Eq.~(\ref{eq:Gamma_result}). In the present section we use these results to
obtain nonperturbative corrections to the thermodynamic quantities at relatively
high temperatures, $G(L_T) \gg 1$. We begin by considering the $N_{ch}\to \infty$
limit in Sec.~\ref{sec:thermo_N_inf} and turn to the case of large, but finite
$N_{ch}$ in Sec.~\ref{sec:thermo_N_fin}.

\subsection{Infinite channel number} \label{sec:thermo_N_inf}

In the $N_{ch}\to \infty$ limit the Coulomb action (\ref{action_c}) vanishes. In
this case  the NL$\sigma$M action has infinitely many degenerate saddle points
with spatially uniform potentials characterized by the winding numbers $w_a$,
$V_a(x)=2\pi T w_a$, with the usual saddle point, $Q=\Lambda$, corresponding to all
$w_a=0$. The single soliton solutions, studied in Sec.~\ref{section:soliton},
represent exact inhomogeneous saddle points with a finite action $G(L_T)$ and
correspond to a kink-like change of of the electric potentials  $V_a(x)$ by $\pm \, 2\pi T$, Eq.~(\ref{eq:saddleV}), in two of the replicas involved in the soliton rotation.
The spatial size of the kink is given by the thermal diffusion length $L_T$. In
the dilute soliton gas limit, which corresponds to  $G(L_T) \gg 1$, multi-soliton
saddle points can be viewed as sets of such kinks separated by distances much
larger than $L_T$. In this case the action of a multi-soliton saddle point is
given by the sum of single soliton actions. Similarly, the functional integral
over the massive modes factorizes into a product of fluctuational determinants
for each soliton. Thus in the dilute regime the soliton gas is noninteracting.
Noting that the sum over the saddle points in Eq.~(\ref{eq:s.p.}) factorizes into
a product of a sum over the uniform saddle points and the sum over the soliton configurations,
we can easily find the multi-soliton contributions to the replicated partition
function in the dilute soliton gas regime,
\begin{equation}\label{eq:s.p.N_inf}
    \langle Z^p \rangle =Z_0^p \sum_{n=0}^\infty \frac{\left[2 p(p-1)\alpha^p\Upsilon_p G(L_T)
    e^{-G(L_T)}
     \right]^n}{n ! }\prod_{i=1}^n \int d\xi^{(i)}_0=Z_0^p
     \exp\left[ 2 p(p-1)\alpha^p\Upsilon_p G(L_T)
    e^{-G(L_T)}L/L_T\right].
\end{equation}
Here $Z_0^p$ denotes the contribution of the homogeneous saddle points to the
replicated partition function, $L/L_T$ is the dimensionless wire length, the
factor of $p(p-1)$ arises from the number of ways the two replicas participating in the
soliton rotation can be chosen from the $p$ replicas available, $\xi^{(i)}_0$ denotes the position of the $i$-th soliton, and the factor of two arises from taking the
Cooperon-like and diffusonlike solitons into account. Substituting this result
into Eq.~(\ref{replicatrick}) we obtain the leading nonperturbative correction to
the average thermodynamic potential in the $N_{ch}\to \infty$ limit:
\begin{equation}\label{eq:deltaF_infty}
\delta \Omega_\infty= 2\Upsilon G(L_T) e^{-G(L_T)} \frac{L}{L_T}T,
\end{equation}
where $\Upsilon$ is defined in Eq.~(\ref{eq:upsilon}).

The correction to the heat capacity can be obtained as $\delta
C_\infty=-T\frac{\partial^2 \delta \Omega_\infty}{\partial T^2}$. Taking into
account that the largest contribution comes from differentiating $G(L_T)$ in the
exponential, we obtain the ratio of $\delta C_\infty$ to the heat capacity of
noninteracting electrons, $C_0=\frac{2\pi^2}{3}\nu A T L$,
\begin{equation}\label{eq:specific_heat_infty}
\frac{\delta C_\infty}{C_0}=-24\Upsilon G^2(L_T)e^{-G(L_T)}.
\end{equation}

The analysis above was restricted to the charge neutrality limit,
$N_{ch}\to \infty$. In the next section we consider the case of a large, but
finite number of channels in the wire. In this case the Coulomb action
(\ref{action_c}) may not be neglected. Its presence significantly  modifies the
behavior of the soliton gas.

\subsection{Finite channel number} \label{sec:thermo_N_fin}

For $N_{ch}\gg 1$ the influence of the Coulomb
action (\ref{action_c}) on the soliton shape and on the massive fluctuations
about the multi-soliton configurations is small and may be neglected. Therefore,
each soliton configuration is still fully
characterized by the soliton positions and the indices of the replicas
participating in the soliton rotation. The Coulomb action for each configuration
is given by the term $S_C$, Eq.~(\ref{action_c}), evaluated for the specific
potential profile $V_a(x)$ corresponding to such a configuration.

For a single soliton situated at $x_0$ the potential profile in the two replicas
participating in the rotation is represented by a kink, $V_0(x-x_0)$,
Eq.~(\ref{eq:saddleV}), in one of the replicas and an antikink, $-V_0(x-x_0)$, in
the other one. Thus each soliton is characterized by its  position $x^{(i)}_0$
and the index of the replica containing the kink, $a_+^{(i)}$, and the antikink,
$a_-^{(i)}$. The potential profile for each soliton configuration is given by
\begin{equation}\label{eq:potential profile}
    V_a(x)=\sum_i \left[ \delta_{a,a_+^{(i)}}- \delta_{a,a_-^{(i)}}
    \right]V_0(x-x_0^{(i)}).
\end{equation}

Using this representation, the replicated partition function, Eq.~(\ref{eq:s.p.}),
can be written as
\begin{equation}\label{eq:s.p.N_fin}
    \langle Z^p \rangle =\tilde{Z}_0^p \sum_{n=0}^\infty
    \frac{\left[2\alpha^p \Upsilon_p G(L_T)
    e^{-G(L_T)}
     \right]^n}{n ! }\prod_{i=1}^n \int d\xi^{(i)}_0 \sum_{a^{(i)}_\pm}
     \exp\left[-S_C\left(\left\{ \xi_0^{(i)},a^{(i)}_\pm\right\}\right)\right],
\end{equation}
where $S_C\left(\left\{ \xi_0^{(i)},a^{(i)}_\pm\right\}\right)$ denotes the
Coulomb action (\ref{action_c}) evaluated for a given soliton configuration
$\left\{ \xi_0^{(i)},a_\pm^{(i)}\right\}$. Since the Coulomb action diverges for
any uniform saddle point with  $w_a\neq 0$, such saddle point are forbidden, and
the factor $\tilde{Z}_0^p$, arising from the uniform saddle points, contains the contribution only from the usual saddle point, $Q=\Lambda$, $\{w_a=0\}$.

Equation (\ref{eq:s.p.N_fin}) is valid in the dilute soliton gas regime,
$G(L_T)\gg1$, in which the typical inter-soliton distance exceeds the thermal
diffusion length $L_T$. In the following we assume that at these length scales
the Coulomb interaction is screened due to the presence of a nearby gate, so that
its Fourier transform is given by $K(q)\approx 1/e^2\ln(\frac{d^2_g}{d^2})$,
where $d_g$ is of the order of the distance to the gate. Then defining the kink
density $\rho_a(\xi)$ in replica $a$,
\begin{equation}\label{eq:kink_density_def}
    \rho_a(\xi) = \sum_i \delta(\xi-\xi^{(i)}_0)\left[\delta_{a,a_+^{(i)}}-
    \delta_{a,a_-^{(i)}}\right],
\end{equation}
we can express the Coulomb action in Eq.~(\ref{eq:s.p.N_fin}) in the dilute gas
limit as
\begin{equation}\label{eq:Coulomb_energy}
    S_C\left(\left\{ \xi_0^{(i)},a_\pm^{(i)}\right\}\right)= -
    \frac{\pi v_F}{32 e^2 \ln\frac{d_g}{d}}  \frac{G(L_T)}{N_{ch}}
    \sum_a \int d\xi d\xi'
     \rho_a(\xi) \rho_a(\xi')|\xi-\xi'|.
\end{equation}

Equations (\ref{eq:s.p.N_fin}), (\ref{eq:kink_density_def}), and
(\ref{eq:Coulomb_energy}) express the replicated partition function of the
disordered wire as a partition function of a one dimensional replicated neutral
gas of kinks and antikinks interacting via a linear potential. Importantly, the
positive and negative charges  in this gas occur only in pairs, such that the
appearance of a positive charge in one replica is accompanied by the appearance
of a negative charge in a different replica at the same spatial position. This
problem can be mapped onto a one-dimensional replicated sine-Gordon
model~\cite{AADPunpublished}. Below we will not use this mapping, but work in the
replicated kink gas representation.

Only the soliton configurations that correspond to a neutral kink gas in each
replica give a nonvanishing contribution to the partition functions because all
non-neutral configurations possess an infinite Coulomb action. The density of the
kink gas is controlled by the fugasity, $\Upsilon_p G(L_T)e^{-G(L_T)}$. Depending
on its  value the kink gas can be in two different regimes. At high temperatures,
for $G(L_T)\gtrsim \ln N_{ch}$, the gas is dimerized. In other words the kinks
within each replica form a dilute gas of bound pairs of a kink and an antikink.
At lower temperatures, $\ln(N_{ch}) \gtrsim G(L_T) \gtrsim1$, the kink pairs
overlap and form an ionized plasma. The dilute soliton gas approximation used to
derive Eq.~(\ref{eq:s.p.N_fin}) is valid in both of these cases. We restrict our
analysis below to the high temperature regime, $G(L_T)> \ln N_{ch}$.

For $G(L_T) \gg 1$, Eq.~(\ref{eq:s.p.N_fin}) may be viewed as an expansion of the
replicated partition function in the powers of the fugasity, $\Upsilon_p
G(L_T)e^{-G(L_T)}$. In the presence of the Coulomb action, the single soliton
contribution to the partition function vanishes, since the corresponding Coulomb
action is infinite. Therefore the leading term in this expansion is given by the
contribution of two solitons which corresponds to two kink-antikink pairs in
different replicas. We shall refer to this object as a dimer.

To evaluate the contribution of a single dimer into the replicated partition
function we express the Coulomb action (\ref{eq:Coulomb_energy}) in terms of the
kink-antikink separation within the dimer, $\xi_{rel}$, and substitute the result
into Eq.~(\ref{eq:s.p.N_fin}). Denoting the dimer center of mass coordinate by
$\xi_{cm}$, summing over all possible pairs of replicas that can accommodate the
dimer, and recalling that each soliton can be either Cooperon-like or
diffusonlike we obtain
\begin{equation}\label{eq:partfnc1}
  \langle
  Z^p\rangle=\tilde{Z}^p_{0}\left(1+4p(p-1)\alpha^p\Upsilon_p^2 G^2(L_T)e^{-2G(L_T)}
  \frac{1}{2!}\int^{L/L_T}_{0}
  d \xi_{cm}\int^{\infty}_{-\infty}d\xi_{rel}e^{-|\xi_{rel}|L_T/L_N}\right),
\end{equation}
where we introduced the notation
\begin{equation}\label{eq:LN}
L_N=\frac{8e^2\ln\frac{d_g}{d}}{\pi v_F}\frac{N_{ch}}{G(L_T)}L_T,
\end{equation}
that has the meaning of the typical kink-antikink separation within each dimer.
Since $L_N \sim L_T/ \sqrt{T \tau_{el}}$, where $\tau_{el}$ is the elastic mean
free time, this length scale is much larger than $L_T$ within the validity domain
of the NL$\sigma$M description. Therefore, the dilute soliton gas approximation is justified. Performing the integrals over $\xi_{cm}$ and
$\xi_{rel}$ in Eq.~(\ref{eq:partfnc1})  we get
\begin{equation}\label{eq:partfnc2}
  \langle
  Z^p\rangle=\tilde{Z}^p_0\left(1+4p(p-1)\alpha^p\Upsilon_p^2 \frac{LL_N}{L^2_T}G^2(L_T)e^{-2G(L_T)}
  \right).
\end{equation}
This expression  shows that the single dimer contribution to the partition
function diverges as the length of the wire $L$ goes to infinity. From the second
term we infer that the spatial density of dimers is $\sim \frac{L_N}{L^2_T}
G^2(L_T) e^{-2G(L_T)}$. In the regime $G(L_T) \gtrsim \ln N_{ch}$ this density is
smaller than $1/L_N$, and multisoliton configurations appear as a dilute gas of
dimers.

Since the dimers in the dilute limit do not interact, the integration over all
dimer configurations results in exponentiation of the correction arising from a
single dimer, second term in Eq.~(\ref{eq:partfnc2}),
\begin{eqnarray}\label{eq:partfnc3}
  \langle Z^p\rangle=\tilde{Z}^p_{0}\left[\sum^{\infty}_{n=0}\frac{1}{n!}\left(4p(p-1)\alpha^p\Upsilon_p^2
  \frac{LL_N}{L^2_T}G^2(L_T)e^{-2G(L_T)}\right)^n\right]=\tilde{Z}^p_{0}e^{4p(p-1)\alpha^p\Upsilon_p^2
 \frac{LL_N}{L^2_T}G^2(L_T)e^{-2G(L_T)}}.
\end{eqnarray}

Using Eq.~(\ref{replicatrick}) and the definition (\ref{eq:LN}) we get  the
expression for the leading nonperturbative correction for the thermodynamic
potential:
\begin{equation}\label{eq:thermodynpot}
  \delta\Omega=\frac{32}{\pi}\Upsilon^2 \frac{e^2}{v_F} \ln\frac{d_g}{d}\,  N_{ch} G(L_T)e^{-2G(L_T)}\frac{L}{L_T}T,
\end{equation}
where $\Upsilon$ is defined in Eq.~(\ref{eq:upsilon}). Using this expression  we
obtain the ratio of the nonperturbative correction to the heat capacity to that
of noninteracting electrons,
\begin{equation}\label{eq:specific heat}
\frac{\delta C}{C_0}= -\frac{384}{\pi}\Upsilon^2 \frac{e^2}{v_F} \ln\frac{d_g}{d}\,
N_{ch} G^2(L_T)e^{-2G(L_T)}.
\end{equation}

Equations~(\ref{eq:thermodynpot}) and ~(\ref{eq:specific heat}) are the main
results of this paper. These results are drastically different from the
expressions (\ref{eq:deltaF_infty}) and (\ref{eq:specific_heat_infty}) obtained
by taking the formal $N_{ch}\to \infty$ limit. We note that the corrections for
the thermodynamic potential for infinite and finite $N_{ch}$,
Eqs.~(\ref{eq:deltaF_infty}) and (\ref{eq:thermodynpot}), become of the same
order at $G(L_T) \sim\ln N_{ch}$, when the dimer gas crosses over into the
ionized regime.

\section{\label{section:Conclusion}Summary}

We studied nonperturbative interaction corrections to the thermodynamic
quantities of a multichannel disordered wire. Within the replica NL$\sigma$M
formalism these corrections arise from soliton saddle points of the NL$\sigma$M
action. In the limit of infinite number of channels, $N_{ch}$,  in the wire we
obtained the exact single soliton solution of the saddle point equations and
evaluated the function integral over the fluctuation about the soliton
configuration. We showed that for $G(L_T)\gg 1$ and $N_{ch} \gg 1$  nonperturbative
corrections to the thermodynamic quantities of the system are described by a
partition function for a dilute gas of solitons. The latter is equivalent to the
partition function for a replicated classical one dimensional Coulomb gas. As the temperature is lowered, this
gas undergoes a crossover from the dimerized regime of neutral soliton pairs at
$G(L_T) > \ln N_{ch}$ to the regime of ionized plasma for $G(L_T) < \ln N_{ch}$.
The crossover $G(L_T) \sim \ln N_{ch} \gg 1$ occurs at temperatures that are
parametrically larger than those corresponding to the transition from weak to
strong localization, $G(L_T) \sim 1$. This enables one to study this crossover
separately from the perturbative effects. We specialized to the high temperature
regime, $G(L_T)\gtrsim \ln N_{ch}$ and obtained the leading nonperturbative
correction to the specific heat (relative to that of noninteracting electrons),
$\delta C/C_0\sim N_{ch} G^2(L_T)e^{-2G(L_T)}$, Eq.~(\ref{eq:specific heat}). We
would like to emphasize that this correction is drastically different from the
result obtained by taking the formal limit $N_{ch}\to \infty$,
Eq.~(\ref{eq:specific_heat_infty}), $\delta C_\infty/C_0\sim
G^2(L_T)e^{-G(L_T)}$.

Although our treatment was specialized to the symplectic ensemble, we believe
that the mapping of the nonperturbative corrections to the soliton gas and to the
replicated Coulomb gas, described by Eqs.~(\ref{eq:s.p.N_fin}) holds for all
three ensembles. Indeed, the existence of soliton minima is generic for all three ensembles~\cite{Andreev}. The mapping to
the replicated classical Coulomb gas relies only on the fact that the functional
integral over the fluctuations about a single soliton configuration can be
reduced to the integral over the soliton position, Eq.~(\ref{eq:Gamma_result}).
This, in turn, is a consequence of the fact that the integral over the massive
modes converges, which we expect to be true for all ensembles.

The generalization of our formalism to the treatment of nonperturbative
corrections to the transport characteristic is left for future work.

\acknowledgements

This work was supported in part by the David and Lucille Packard Foundation.

\appendix

\section{\label{section:expression for det}Derivation of Eq.~(\ref{eq:exprgamma_h})}

In Sec.~\ref{section:integration_W} we encountered expressions containing
determinants of Schr\"{o}dinger-type operators of the form
\begin{subequations}
\begin{eqnarray}
\label{eq:localdet}
  \ln{D_0}&=&\textrm{tr}_\xi\ln\frac{\omega-\sder{}{\xi}+U}{\omega-\sder{}{\xi}},\\
\label{eq:steplikedet}
\ln{D_h}&=&\tr_\xi\ln\frac{\left(\omega-\sder{}{\xi}+U_1\right)
  \left(\omega-\sder{}{\xi}+U_2\right)}
  {\left(\omega-\sder{}{\xi}+h\right)\left(\omega-\sder{}{\xi}\right)},
\end{eqnarray}
\end{subequations}
where $U(\xi)$ is a potential that vanishes at
$\xi\rightarrow\pm\infty$, and $U_{1,2}(\xi)$ are step-like potentials,
satisfying $U_1(\xi)=U_2(-\xi)$, $U_1(-\infty)=0$, $U_1(\infty)= h$.

As explained in the text above Eq.~(\ref{eq:exprgamma_0}), the trace
in~(\ref{eq:localdet}) can be obtained as a particular case of that
in~(\ref{eq:steplikedet}). Therefore, we concentrate our attention on the latter.
We first rewrite Eq.~(\ref{eq:steplikedet}) as
\begin{eqnarray}\label{eq:steplikedet1}
\ln{D_h}&=&\tr_\xi\ln\frac{\omega-\sder{}{\xi}+U_1}{\omega-\sder{}{\xi}+h\Theta}+
  \tr_\xi\ln\frac{\omega-\sder{}{\xi}+U_2}{\omega-\sder{}{\xi}+h(1-\Theta)}+
  \tr_\xi\ln\frac{\left(\omega-\sder{}{\xi}+h\Theta\right)\left(\omega-\sder{}{\xi}+h(1-\Theta)\right)}
{\left(\omega-\sder{}{\xi}+h\right)\left(\omega-\sder{}{\xi}\right)},
\end{eqnarray}
where $\Theta(\xi)$ is the step function. The third term does not depend on the
potentials, and can be calculated explicitly, which is done at the end of this
Appendix. The first two terms are equal since $U_1(\xi)=U_2(-\xi)$. We denote
each of them as $\ln D_{h1}$, and proceed to calculate this quantity.

To compute
\begin{equation}
\ln
D_{h1}=\tr_\xi\ln\frac{\omega-\sder{}{\xi}+U_1}{\omega-\sder{}{\xi}+h\Theta},
\end{equation}
we represent the potential $U_1(\xi)$ as a sum $U_1(\xi)=h\Theta(\xi)+v(\xi)$,
where $v(\xi)$ vanishes at spatial infinities, and express the variational
derivative of $\ln D_{h1}$ with respect to $v(\xi)$ in terms of the Green's
function $G(\xi,\xi')=\left.\left(\omega-\sder{}{\xi}+U_1\right)^{-1}\right|
_{\xi, \xi'}$,
\begin{eqnarray}\label{eq:funcder}
  \frac{\delta \ln{D_{h1}}}{\delta v(\xi)}=\frac{\delta \tr_\xi\ln G^{-1}}{\delta
  v(\xi)}=G(\xi,\xi).
\end{eqnarray}
The Green's function $G(\xi,\xi')$ can be found by solving the differential
equation
\begin{eqnarray}
  \left[\omega-\sder{}{\xi}+U_1(\xi)\right]G(\xi,\xi')=\delta(\xi-\xi'),
\end{eqnarray}
with the boundary conditions that $G(\xi,\xi')$ vanishes at spatial
infinities $\xi, \xi' \to \pm \infty$. It can be expressed~\cite{mathref} in terms
of the two independent solutions of the homogeneous equation
\begin{equation}\label{eq:SE}
\left[\omega-\sder{}{\xi}+U_1(\xi)\right]\phi_i(\xi)=0,
\end{equation}
such that $\phi_1(\xi \to-\infty) \to 0$ and $\phi_2(\xi \to
+\infty) \to 0$. In particular, at coinciding points we have
\begin{equation}\label{eq:gfnonloc}
  G(\xi,\xi)=\frac{\phi_1(\xi)\phi_2(\xi)}{W[\phi_1(\xi),\phi_2(\xi)]},
\end{equation}
where $W[\phi_1(\xi),\phi_2(\xi)]$ is the Wronskian of $\phi_1(\xi)$
and $\phi_2(\xi)$,
\begin{equation}
  W[\phi_1(\xi),\phi_2(\xi)]=\der{\phi_1(\xi)}{\xi}\phi_2(\xi)
  -\phi_1(\xi)\der{\phi_2(\xi)}{\xi}.
\end{equation}
The Wronskian of the two independent solutions of
Eq.~(\ref{eq:SE}) does not depend on coordinate $\xi$, and therefore
may be expressed in terms of the $\xi \to \pm \infty$ asymptotics of
$\phi_i(\xi)$. By appropriately normalizing the solutions, we can express the latter as
\begin{eqnarray}\label{boundarycondnonloc}
  \phi_1(\xi)=\left\{
\begin{tabular}{l}
 $e^{k\xi}, \, \xi\rightarrow -\infty$\\
 $te^{\varsigma \xi}, \, \xi\rightarrow\infty$
\end{tabular}
  \right., \quad
\phi_2(\xi)=\left\{
\begin{tabular}{l}
$t'e^{-k\xi},\,  \xi\rightarrow -\infty$\\
 $e^{-\varsigma \xi},\, \xi\rightarrow \infty$
 \end{tabular}
  \right.,
\end{eqnarray}
where $k=\sqrt{\omega}$, $\varsigma=\sqrt{\w+h}$, and $t$, $t'$
depend on the specific form of the operator in Eq.~(\ref{eq:SE}).
Evaluating the Wronskian at $\xi\rightarrow\pm\infty$ using the
asymptotics (\ref{boundarycondnonloc}), we obtain
\begin{equation}\label{eq:W_t}
    W[\phi_1(\xi),\phi_2(\xi)]=2\varsigma t=2kt'.
\end{equation}

Next we prove that
\begin{equation}\label{eq:varder}
 \frac{\delta \ln{D_{h1}}}{\delta v(\xi)}=\frac{\delta \ln{t}}{\delta
 v(\xi)}.
\end{equation}
To this end we introduce an auxiliary construction
\begin{equation}
  \tilde{W}[\phi_1(\xi),\tilde{\phi}_2(\xi)]= \der{\phi_1(\xi)}{\xi}\tilde{\phi}_2(\xi)
  -\phi_1(\xi)\der{\tilde{\phi}_2(\xi)}{\xi}.
\end{equation}
Here $\phi_1$ and $\tilde{\phi}_2$ are solutions of
Eq.~(\ref{eq:SE}) with the same $\omega$, but for two different
potentials $v(\xi)$ and $\tilde{v}(\xi)$, both of which vanish at
$\xi\rightarrow\pm\infty$. The tilde denotes quantities
corresponding to $\tilde{v}$. We assume that $\phi_1$ and
$\tilde{\phi}_2$ have the asymptotic form (\ref{boundarycondnonloc}),
with $\tilde{\phi}_2$ characterized by $\tilde{t}'$.

In contrast to the Wronskian $W[\phi_1(\xi),\phi_2(\xi)]$, built out of
the solutions of the same equation, the quantity
$\tilde{W}[\phi_1(\xi),\tilde{\phi}_2(\xi)]$ depends on the
coordinate and satisfies the differential equation
\begin{equation}\label{eq:eomwronsk}
\frac{d\tilde{W}(\xi)}{d\xi}=\sder{\phi_1(\xi)}{\xi}\tilde{\phi}_2(\xi)
  -\phi_1(\xi)\sder{\tilde{\phi}_2(\xi)}{\xi}=
  \left[v(\xi)-\tilde{v}(\xi)\right]\phi_1\tilde{\phi}_2,
\end{equation}
that follows directly from Eqs.~(\ref{eq:SE}) for $\phi_1$ and
$\tilde{\phi}_2$.

Integrating Eq.~(\ref{eq:eomwronsk}) with respect to $\xi$ from
$-\infty$ to $\infty$ and using the asymptotic form of $\phi_1$ and
$\tilde{\phi}_2$, Eq.~(\ref{boundarycondnonloc}), we obtain
  \begin{equation}
    \tilde{W}(\infty)-\tilde{W}(-\infty)=
    \int^{\infty}_{-\infty}d\xi(v-\tilde{v})\phi_1\tilde{\phi}_2=2\varsigma t-2k\tilde{t}'.
  \end{equation}
Taking a variational derivative of this equation with respect to
$v(\xi)$ at $v(\xi)=\tilde{v}(\xi)$ we obtain
\begin{equation}\label{eq:var_deriv_t}
    \phi_1(\xi)\phi_2(\xi)=2\varsigma\frac{\delta t}{\delta
    v(\xi)}.
\end{equation}
Plugging Eqs.~(\ref{eq:W_t}) and~(\ref{eq:var_deriv_t}) into
Eq.~(\ref{eq:gfnonloc})  for the Green's function, and using
Eq.~(\ref{eq:funcder}), we obtain Eq.~(\ref{eq:varder}).

Integrating Eq.~(\ref{eq:varder}) with respect to $v$ from
$v(\xi)=0$ to its final value we obtain
\begin{equation}\label{eq:d1h} \ln D_{h1}=\ln \frac{t}{t_0},
\end{equation}
where $t_0$ is the coefficient in front of $e^{\varsigma \xi}$ in
asymptotic form (\ref{boundarycondnonloc}) of $\phi_1$ for
$v(\xi)=0$. The latter can be easily found from the continuity of
the logarithmic derivative $d \ln \phi_1(\xi)/d\xi$ at $\xi=0$, and is
given by $t_0=(1+k/\varsigma)/2$.

Finally, the third term in Eq.~(\ref{eq:steplikedet1}) can be
calculated in the following manner. We denote this term by
$T_3(\w)$, and introduce the Green's functions $g_0$, $g_h$, and $g^{\pm}$  that
vanish at $\xi, \xi' \to \pm \infty$  and satisfy the equations
\begin{eqnarray}\label{eq:g}
  \left(\w-\sder{}{\xi}\right)g_0&=&\delta(\xi-\xi'),\,\,\,
  \left(\w-\sder{}{\xi}+h\right)g_h=\delta(\xi-\xi'),\nonumber\\
  \left(\w-\sder{}{\xi}+h\Theta\right)g^{+}&=&\delta(\xi-\xi'),\,\,\,
  \left(\w-\sder{}{\xi}+h(1-\Theta)\right)g^{-}=\delta(\xi-\xi').
\end{eqnarray}

Taking the derivative of $T_3(\w)$  with respect to $\w$ we obtain
\begin{equation}\label{eq:T3}
\pder{T_3}{\w}=\int_{-\infty}^\infty d \xi \, [g^+(\xi,\xi)
+g^-(\xi,\xi)-g_0(\xi,\xi)-g_h(\xi,\xi)].
\end{equation}
All Green's functions entering this equation are easily calculated
using the method of Wronskian, as was done above for a general
potential. Specifically, we obtain the following expressions for the
Green's function at coinciding points:
\begin{eqnarray}
&&g_0(\xi,\xi)=\frac{1}{2k},\,\,\,g_h=\frac{1}{2\varsigma},\,
\,\,g^-=g^+(-\xi,-\xi'),\nonumber\\
&&g^+(\xi,\xi)=\Theta(-\xi)
\left(\frac{1}{2k}+\frac{k-\varsigma}{2k(k+\varsigma)}e^{2k\xi}\right)+
\Theta(\xi)\left(\frac{1}{2\varsigma}+
\frac{\varsigma-k}{2\varsigma(k+\varsigma)}e^{-2\varsigma\xi}\right).
\end{eqnarray}
Keeping in mind that $T_3(\w \to \infty) \to 0$, we can express it as
$T_3(\w)=-\int^\infty_{\w} d\w'\pder{T_3(\w')}{\w'}$. Substituting
the expressions~(\ref{eq:g}) for the Green's functions into Eq.~(\ref{eq:T3}), we obtain
\begin{equation}
\label{eq:T3_result}
  T_3=\ln{\frac{\varsigma t^2_0}{k}},
\end{equation}
with $t_0$ defined below Eq.~(\ref{eq:d1h}).

Substituting Eqs.~(\ref{eq:T3_result}) and (\ref{eq:d1h}) into
Eq.~(\ref{eq:steplikedet1}) we obtain the final expression for trace
in Eq.~(\ref{eq:steplikedet}),
\begin{equation}\label{eq:detnonlocal}
  \tr_\xi\ln\frac{\left(\omega-\sder{}{\xi}+U_1\right)
  \left(\omega-\sder{}{\xi}+U_2\right)}
  {\left(\omega-\sder{}{\xi}+h\right)\left(\omega-\sder{}{\xi}\right)}=
  2\ln D_{h1}+T_3=\ln{\sqrt{\frac{\w+h}{\w}}t^2}=\ln{tt'},
\end{equation}
where in the last expression we used $\sqrt{\w+h} t=\sqrt{\w}t'$ to
write a more symmetric expression, in which $t$ and $t'$ are defined
in~(\ref{boundarycondnonloc}). This proves
Eq.~(\ref{eq:exprgamma_h}).

\end{document}